# All-optical control of second-harmonic generation in β-BaB$_2$O$_4$ via coherent, terahertz-driven acentric lattice displacement

Flavio Giorgianni[1,2,*], Nicola Colonna[2], Gabriel Nagamine[1], Leonie Spitz[2], Guy Matmon[2], Alexandre Trisorio[2], Nicolas Forget[3], Carlo Vicario[2], Adrian L. Cavalieri[1,2]

[1]*Institute of Applied Physics, University of Bern, CH-3012 Bern, Switzerland*

[2]*Paul Scherrer Institute, CH-5232 Villigen-PSI, Switzerland*

[3]*Institut de Physique de Nice (INPHYNI), Université Côte d'Azur, France*

Dynamical control of the nonlinear optical properties of solids – with light itself – will be essential for future ultrafast photonic technologies. Previously, methods to modulate nonlinear processes including second-harmonic generation (SHG) have relied primarily on non-resonant light-matter interaction or photo-generation of hot electrons in nanoscale materials. However, these approaches are typically constrained by limited interaction lengths and the initial frequency conversion is relatively weak under equilibrium conditions. Here, a ~30% modulation of efficient phase-matched SHG in bulk beta-barium borate (β-BaB$_2$O$_4$) is achieved through transient lattice deformation by intense terahertz (THz) pulses that are tuned to resonance with an infrared-active phonon mode. The effect originates from modification of the index of refraction ellipsiod and the corresponding nonlinear phase-matching conditions, rather than from direct modulation of the nonlinear susceptibility through THz-mediated $\chi^{(3)}$ processes. This mechanism, of resonant selective lattice excitation, points toward novel THz-control schemes to tune the nonlinear optical response in materials.

## Introduction

Nonlinear optical frequency conversion – including second harmonic generation (SHG) – greatly extends the spectral range of conventional lasers, underpinning key scientific applications across a wide range of fields from high-resolution microscopy to precision optical metrology to quantum information science[1–6]. For communications and technological applications, integrated photonic circuits have emerged[7–10], in which the nonlinear optical response can be precisely modulated. Control over optical nonlinearities, including SHG, using applied electric fields has now been demonstrated in optical crystals[11], polymers[12], semiconductors[13–15], atomic-layer systems and heterostructures,[16,17] and artificial photonic structures[18,19]. However, modulation of the nonlinear optical response with this approach is fundamentally limited by the speed of conventional electronics to GHz modulation rates, creating a functional barrier to the next generation of optoelectronic devices. In principle, orders-of-magnitude higher speeds can be achieved through all-optical approaches, which would enable ultrafast data transfer and real-time optical signal processing.

---

[*]Corresponding Author
E-mail: Flavio.giorgianni@unibe.ch

Indeed, ultrafast, all-optical SHG modulation has been demonstrated in nanostructures and two-dimensional materials[20–23]. However, these achievements appear impractical for applications in solid-state devices, as the modulation generally relies on either photoexcitation of hot electrons[22,24] with significant collateral heating, or non-resonant light-matter interaction,[25–29] which is limited by material damage thresholds. Further, while high contrast has been achieved, the absolute SHG signals themselves remain relatively weak due to limited interaction lengths. On the other hand, while bulk three-dimensional materials can deliver strong SHG signals suitable for applications, high-contrast, ultrafast, all-optical modulation has not yet been reported.

In this work, high-contrast modulation of strong SHG is achieved by resonant excitation of the β-barium borate (β-BaB$_2$O$_4$, BBO) crystal lattice using intense terahertz (THz) pulses tuned to an infrared (IR)-active phonon mode. Theoretical considerations — based on group theory and *ab-initio* calculations[30–32] — indicate that both the linear properties and the second-order optical nonlinearity in BBO originates from localized electronic density within the B$_3$O$_6$ anionic rings in the crystal unit cell. Therefore, displacement of the anionic ring can be predicted to significantly influence SHG and targeted displacements might lead to strong, controlled modulation of the SHG signal. While this parameter cannot be selectively accessed with conventional stimuli (e.g., strain, static electric fields), selective displacement can be achieved with intense THz pulses, resonantly coupled to IR-active phonons, as illustrated in Fig. 1a. With this approach, THz pulses can be expected to enable ultrafast control over the linear and nonlinear optical properties of BBO, enabling efficient modulation of the SHG process.

The potential for control by THz-driven resonant excitation is investigated here through the SHG of a femtosecond NIR probe laser. By measuring the time-resolved SHG efficiency as a function of the fundamental input wave polarization and delay with respect to the THz excitation, the modulated SHG signal yields the overall out-of-equilibrium nonlinear optical response. And importantly, under appropriate input conditions, high-contrast SHG modulation exceeding 30% can be reached.

Macroscopically, this effect can be explained by an electro-optic, THz-induced rotation of the principal optical axes (perpendicular to the direction of propagation), which alters the SHG phase matching conditions. Density functional perturbation theory (DFPT) calculations validated by the experiment can then be used to obtain an accurate estimate of the frequency dispersion of the electro-optic coefficient in the THz range, as well as the phonon normal-mode amplitude of 0.67 Å(amu)$^{1/2}$, which displaces the B$_3$O$_6$ anionic unit from its equilibrium position. This quantitative link between the atomic-scale structural dynamics and macroscopic nonlinear optical response may facilitate engineering of materials with specific transient nonlinear optical responses in the future.

## Results:

## Experimental design and THz excitation geometry

To investigate the SHG modulation by selective lattice excitation on ultrafast timescales, intense THz pulses are used to drive the material while its linear and nonlinear response are probed with an overlapping femtosecond near-infrared (NIR) pulse (Fig. 1a-b and Secs.1 and 2 of the Supplementary Information for details). Prior to probing the dynamics, the BBO crystal is characterized in the chosen experimental geometry in which the THz pulse is polarized along the ordinary axis (o-axis) of the BBO crystal. Steady-state optical spectroscopy (see Supplementary Fig. 3 and Sec. 3 of the Supplementary Information for details) yields the imaginary part of the dielectric function $\varepsilon_2(\omega)$ and the corresponding IR-active phonon modes (Fig 1d). The dominant IR-active $E$-symmetry phonon mode is found at a frequency of $\omega_Q = 4.32$ THz, consistent with other recent independent characterizations by THz time-domain spectroscopy[33]. This mode is of particular interest, as density functional perturbation theory (DFPT) calculations indicate that it leads to significant co-planar motion of anionic $[B_3O_6]^{3-}$ rings while the $Ba^{2+}$ cations will exhibit in nearly motion in the opposite direction with comparatively smaller displacements, due to their large atomic mass[31] (Fig. 1a).

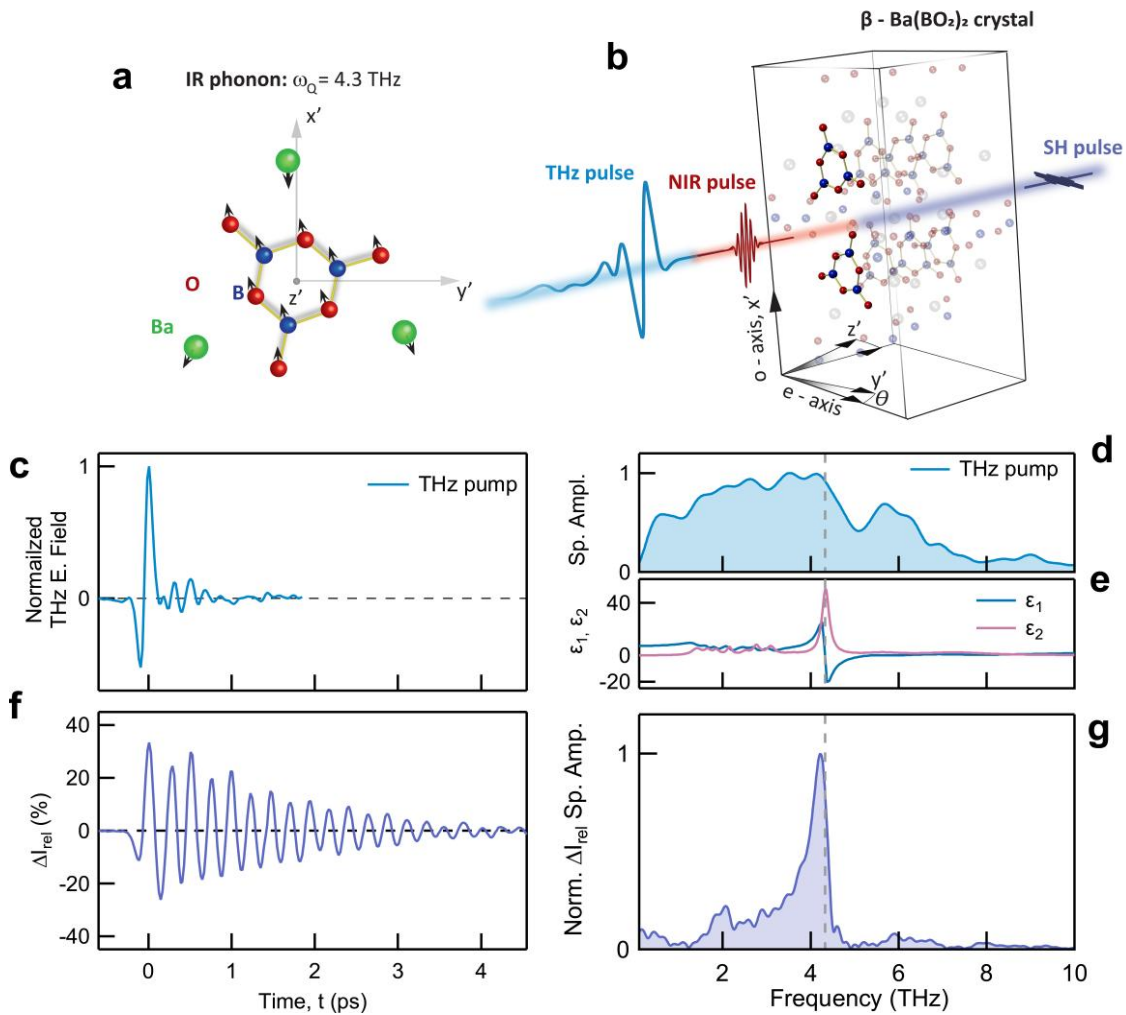

*Fig. 1: THz-driven optical SHG modulation dynamics of β-BaB₂O₄:* **a**, *Illustration of the planar $B_3O_6$ anionic unit surrounded by Ba atoms within the unit cell of β-BaB₂O₄ in the crystallographic coordinates x' and y', along with the*

*phonon displacement vectors projected in the crystallographic x'-y' plane associated with the IR-active mode at $\omega_Q = 4.3\ THz$. Phonon displacement vectors have been calculated using density functional theory. **b**, Schematic of the experimental setup. The THz pulse interacts with the nonlinear crystal while the SHG response is probed with a temporally synchronized ~50 femtosecond NIR pulse (FF beam) converted in a SH pulse. Arrows define the orientation of the ordinary and extraodinary axes with respect to the crystallographic axes x', y', and c ($\theta = 29.3°$). **c**, Temporal waveform of the THz electric field and **d**, corresponding normalized Fourier spectral amplitude (Sp. Ampl.). **e**, Real and imaginary components ($\varepsilon_1, \varepsilon_2$) of the dielectric function of BBO (β-BaB$_2$O$_4$) measure using FTIR spectroscopy. **f**, THz-driven SH intensity mo$\varepsilon_2$ dulation dynamics (blue curve). **g**, Normalized Fourier spectral amplitude (Sp. Ampl.) of the SH harmonic modulation dynamics is shown in panel **e**.*

Electro-optic sampling (EOS) of the driving THz pulse is used to verify spectral overlap with the $E$ phonon mode required for targeted excitation. In-situ EOS measurements of the THz temporal waveform $E_{THz}(t)$ yield a peak electric field strength of $E_{THz}^{peak} = 8.5$ MV/cm, and Fourier analysis reveals a spectrum heavily weighted around 4.3 THz, with additional weaker frequency components ranging from $0 - 10$ THz (Fig. 1c-d). The spectral content ensures efficient coupling of the THz pulse to the desired $E$ phonon mode. While there are additional IR-active phonon modes spanned by the broadband THz pulse, they have lower absorption cross-sections[33] and are not expected to contribute significantly to the driven lattice dynamics (Fig. 1e).

Following initial characterization, time-resolved experiments are performed by focusing the THz pulse into a 300 μm thick BBO crystal cut for type-I phase-matched SHG of a ~50 femtosecond NIR laser pulse at the fundamental wavelength of 800 nm. In the experiment, 20 nJ NIR pulses are focused to a spot size of 35 μm at the BBO crystal with corresponding intensity of ~ $8 \times 10^{10}$ W/cm$^2$, and at equilibrium a maximum SHG energy conversion efficiency of ~ 4 % is observed.

By varying the delay between the THz pump pulse and polarization of the femtosecond NIR probe pulse, the influence of the phonon dynamics on the SHG response is investigated in the time domain as function of polarization. A key observable is the time-dependent, normalized relative strength of the SHG intensity, $\Delta I_{rel} = \Delta I_{SH}/I_{SH,0}$. Here, $\Delta I_{SH} = I_{SH,THz} - I_{SH,0}$, $I_{SH,THz}$, and the equilibrium SHG intensity, $I_{SH,0}$. By convention, time-zero occurs when the peak of the femtosecond NIR pulse coincides with the maximum of the THz electric field. Under optimized input polarization (detuned from peak conversion efficiency) and delay the THz pulse induces a modulation in the relative SHG strength exceeding 30% (Fig. 1f), whereas no significant THz-induced modulation is observed in the NIR probe intensity (Supplementary Fig. 2).

The time-dependent SHG exhibits a damped oscillatory behavior persisting for over 4 ps, with Fourier analysis revealing a distinct peak slighlty below the phonon frequency, $\omega_Q$, coincident with the positive maximum of real part of the dielectric fuction, $\varepsilon_1$, see Fig. 1g. Tuning the spectrum of the THz driving field with spectral filters to above and below the phonon frequency $\omega_Q$. When the spectral components of the driving THz field are above $\omega_Q$, the SHG modulation is suppressed. Similarly, when the THz cut-off frequency is below $\omega_Q$, modulation of the SHG is also strongly suppressed and can be attributed to weak excitation of lower frequency phonon modes (Fig. 2a).

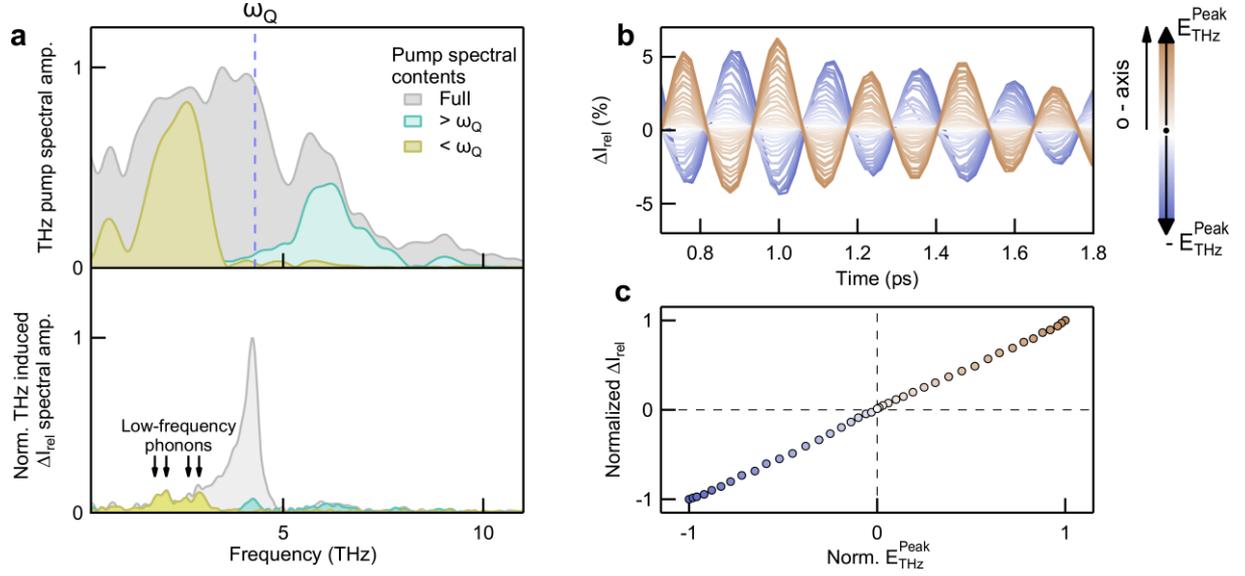

*Fig. 2: Phonon-induced SH intensity modulation dynamics of β-BaB₂O₄. a, THz excitation of the BBO IR-active phonon modes. On resonance, a pronounced spectral amplitude (amp.) response is observed (grey curves). For comparison, low and high frequency off-resonant THz excitation (yellow and green curves) leads to a significantly reduced modulation of the second harmonic intensity. b-c, The amplitude of the SH intensity modulation is directly proportional to the resonant peak field $E_{THz}^{peak}$ of the THz pump and is inverted upon reversing the polarity of the applied THz field.*

The phase dependence of the SHG modulation is further investigated by returning to the nominal THz pulse (tuned to resonance) and inverting its polarity. A corresponding inversion in the relative SHG intensity is observed, indicating that the dynamics are resonantly driven and phase-locked to the THz field (Fig. 2b). Additionally, scaling the THz field strength results in a linearly dependent effect, suggesting that the SHG modulation is directly proportional to the phonon amplitude, which is itself linearly dependent on the THz driving field strength[34–36] (Fig. 2c).

**Polarisation dependence**

At equilibrium, the expression for the type-I SHG in the undepleted pump approximation in BBO of length $L$ by an electromagnetic wave of intensity $I$, frequency $\omega$, and polarisation angle $\alpha$ is given by (see Supplementary Information Sec. 4):

$$I_{SH,0}(\alpha) = \frac{2\omega^2 d_{NL}^2 L^2}{n_e n_o^2 c^3 \epsilon_0} \left( \frac{\sin\left(\frac{1}{2}\Delta k\, L\right)}{\frac{1}{2}\Delta k\, L} \right)^2 \cos^4(\alpha)\, I^2(\omega), \tag{1}$$

where $n_o(\omega)$ is the ordinary refractive index at the fundamental ($\omega$) and $n_e(2\omega)$ is the extraordinary refractive index at second harmonic ($2\omega$), $d_{NL}$ is the nonlinear optical coefficient, and $\Delta k$ represents the momentum mismatch between the fundamental and second harmonic waves ($\Delta k = 0$ in perfect phase matching conditions). The polarisation angle $\alpha$ defines the orientation of the fundamental wave with respect to the ordinary axis, as illustrated in Fig. 3a. According to Eq. (1), and consistent with equilibrium experimental observations, the SHG intensity $I_{SH,0}(\alpha)$ exhibits a cosine-to-the-fourth power dependence on input polarisation, resulting in a two-fold symmetric angular distribution with maximum intensity at $\alpha = 0$ and $\alpha = 180°$ (Fig. 3b).

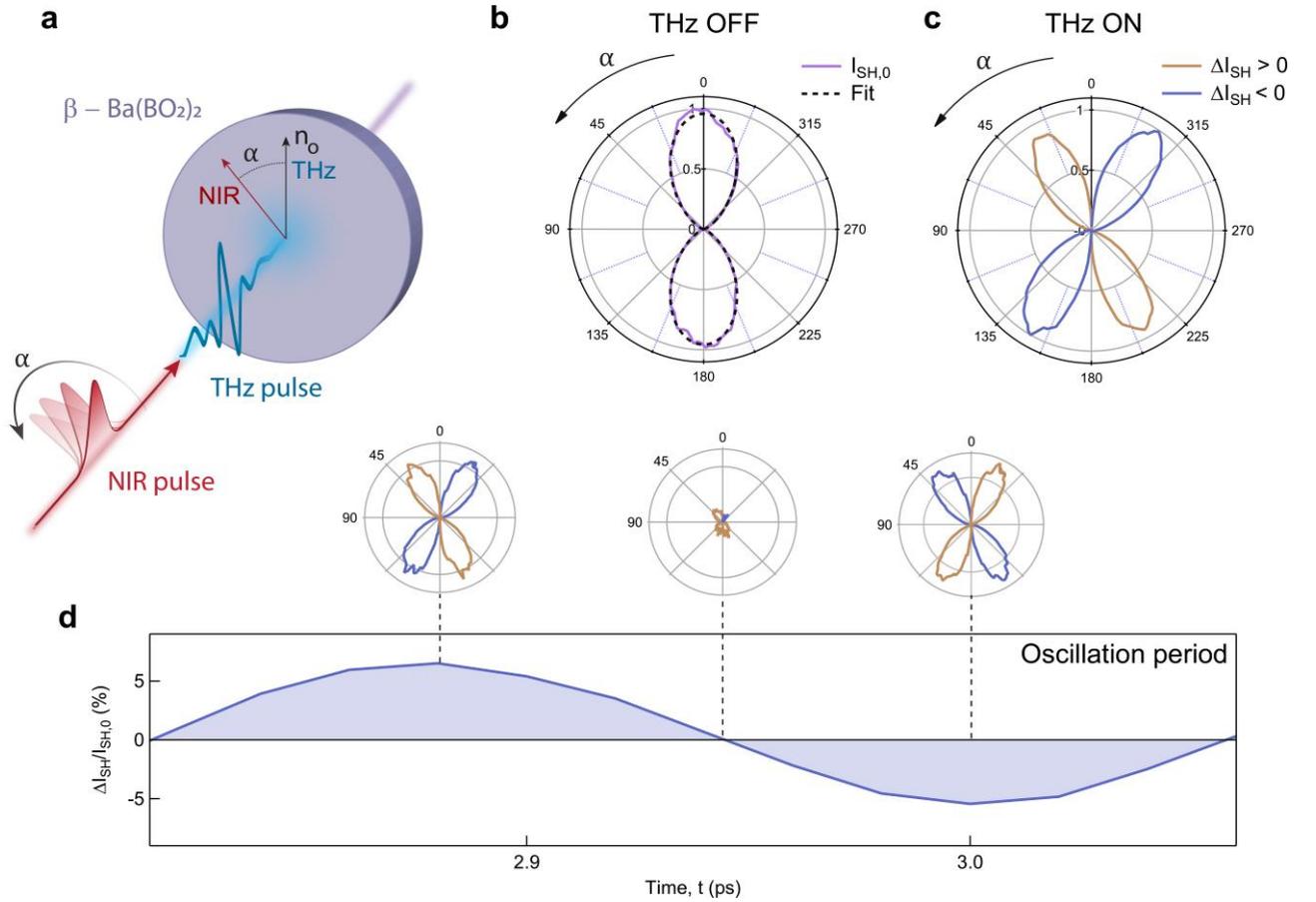

*Fig. 3. Polarization resolved THz-induced SHG modulation: a. The angle $\alpha$ between the NIR pulse polarization and the BBO ordinary principal axis is varied while the THz field is kept parallel to the ordinary axis. b. The SHG intensity as a function of $\alpha$ is fit by $\cos^4 \alpha$ curve (black dashed line), which is the expected angular dependence in type-I SHG. c. THz-induced SHG modulation $\Delta I_{SH}(\alpha)$ versus $\alpha$ at fixed THz-NIR delay time of 2.4 ps. The orange solid curve denotes an increase of the SH intensity ($\Delta I_{SH}(\alpha) > 0$) while the blue line a reduction ($\Delta I_{SH}(\alpha) < 0$). d. The positive and negative modulation quadrants are interchanged by delaying the THz-NIR by one phonon half period.*

Out-of-equilibrium, the SHG can in principal be modulated by two distinct effects. The dynamics can result from a time-dependent variation in the second-order nonlinear coefficient, which is proportional to the driving THz field and can be considered effectively as a third-order $\chi^{(3)}$ effect. Notably, under certain circumstances, this $\chi^{(3)}$-driven effect has been used to describe THz-Field Induced Second Harmonic Generation (TFISH) [37–40]. Alternatively, the dynamics can result from phonon-mediated electro-optic effects in the linear index of refraction (a $\chi^{(2)}$ effect involving THz and phonon frequencies) – and subsequent modification of the phase-matching conditions ($\Delta k$), affecting the SHG efficiency, without inducing a time-dependent $\chi^{(2)}$ or $\chi^{(3)}$ effect, *through the second-order nonlinear coefficient involving the fundamental field frequency* (Secs. 5 and 6 of the Supplementary Information). This mechanism has been described as a cascade of $\chi^{(2)}$ processes[26,41]. Under strong-field THz excitation, both $\chi^{(2)}$ and $\chi^{(3)}$ effects are always present, however, it can be expected that only one of the effects dominates the nonlinear response depending on the specific equilibrium phase-matching conditions.

To identify the origin of the observed dynamics, the NIR probe pulse polarization is scanned at a time delay of $t = 2.4$ ps - corresponding to a local maximum of the SHG dynamics. Notability, the angular dependence of the measured SHG modulation, $\Delta I_{SH}(\alpha) = I_{SH,THz}(\alpha) - I_{SH,0}(\alpha)$, now exhibits a four-fold symmetry, with the two lobes peaking at $\alpha \approx 30°$ and $\alpha \approx 330°$ having a positive modulation ($\Delta I_{SH}(\alpha) > 0$), corresponding to an increase in SHG intensity (Fig 3) and vice-versa for the lobes observed at $\alpha \approx 150°$ and $\alpha \approx 210°$. Further, as the temporal delay between the THz and NIR pulses is scanned over one oscillation cycle, the sign of $\Delta I_{SH}(\alpha)$ is inverted in all lobes, passing through a common zero-crossing where no modulation is observed ($\Delta I_{SH}(\alpha) = 0$), as shown in Fig. 3d.

Importantly, this angular and temporal dependence of the SHG signal in the type-I phase matched geometry is incompatible with a TFISH interpretation involving an effective $\chi^{(3)}$-effect resulting from a THz-induced modulation of the second-order $\chi^{(2)}$ susceptibility. Evidence of this inconsistency is obtained from comparative measurements of SHG modulation in BBO where the phase-matching conditions are not met (see Supplementary Fig. 5). Here, due to the limited coherence length, the the $\chi^{(3)}$ effects to be isolated, as electro-optically induced changes in the index of refraction ellipsoid are negligible in the overall SHG response. In the phase-mismatched conditions, the $\chi^{(3)}$-effects dominate, resulting in a polarization dependence in the SHG modulation with features fundamentally different from those observed in the phase matched case. Additionally, the equilibrium signal strength and magnitude of the modulated intensity is orders of magnitude weaker (see Secs. 7 and 8 of the Supplementary Information for full explanation). Therefore, in the current experiment, the strength of the effect and the 4-fold symmetry leads to the conclusion that the dynamic effect is dominated by the phonon-mediated alteration of the type I phase-matching conditions – a cascaded 2$^{nd}$-order effect.

**Analytical model of the cascaded $\chi^{(2)}$ dynamics**

Resonant excitation of the E phonon mode by the THz pulse displaces the $(B_3O_6)^{3-}$ anionic rings far from equilibrium positions (Fig 1a), thereby inducing an electro-optic modulation of the refractive index along the crystal's principal axes $(x', y', z';$ see Fig. 1a). Under equilibrium conditions, BBO is a negative uniaxial crystal $(n_{x'} = n_{y'} \neq n_{z'})$, but the phonon excitation breaks its axial symmetry, leading to biaxial behaviour where all three refractive indices become distinct $(n_x \neq n_y \neq n_z)$ – see Sec. 5 of the Supplementary Information for details. The dynamical refractive indices along the $x$ and $y$ axes in the crystal-frame become: $n_{x'}^{THz} = n_{x'} + \delta n$ and $n_{y'}^{THz} = n_{y'} - \delta n$, where the magnitude of the phonon-induced refractive index modulation is:

$$\delta n = n_{x',y'}^3 r_{22} Q/2 \qquad (2)$$

Here, $Q$ is the phonon amplitude and $r_{22}$ is the phonon-optic coupling coefficient, relating the modulation of the refractive index to the phonon amplitude. To interpret the index of refraction ellipsoid in the context of phase matching, which governs the efficiency of the SHG $\chi^{(2)}(2\omega, \omega, \omega)$-process, a projection normal to the direction of propagation in the laboratory frame is taken from which the indices along the ordinary and extraordinary axes, known as the principal axes, are found.

In the experiment, the BBO crystal was cut for optimal type-I phase matching at equilibrium (SHG at 800 nm), with its crystal axes oriented at θ = 29.3° and φ = 90° (following the conventions in Ref. [42]). The cut angle defines the direction of propagation in the crystal system and ensures that the refractive index of the ordinary fundamental wave $(n_o)$ matches that of the extraordinary second harmonic wave $(n_e)$. In the dynamical case, the phonon-induced modulation of the index of refraction ellipsoid ($\delta n$ in the crystal coordinate system) leads to a rotation of the ordinary and extraordinary axes in the laboratory frame by an angle $\gamma = \kappa(\theta) r_{22} Q$, as illustrated in Fig. 4a. Here, $\kappa(\theta)$ is a geometrical parameter that depends on the crystal orientation $\theta$.

While the resonant phonon excitation rotates the ordinary and extraordinary axes, impacting the phase-matching conditions in the SHG process, the magnitudes of the ordinary and extraordinary refractive indices – apart from a third-order correction – remain nearly unchanged. (Derivations of the modulation in the refractive index $\delta n$, the effective rotation angle $\gamma$ of the principal axes, and all relevant details are given in Secs. 5 and 6 of Supplementary Information).

To quantitatively model the dynamics, it is also essential to consider the spatio-temporal propagation of the phonon inside the BBO crystal. The phonon amplitude decays exponentially along the direction of propagation. This spatial dependence of the phonon amplitude can be described by $Q(z) = Q_0 e^{-\frac{z}{\delta_{ph}}}$, where $Q_0$ is the initial phonon amplitude at the crystal surface, and $\delta_{ph}$ is its penetration depth. By accounting for the phonon-induced rotation of the ordinary and extraordinary optical axes, as well as the penetration depth of the phonon, the variation in SHG intensity as a function of the input polarization angle $\alpha$ (at fixed THz-NIR delay) is the given by:

$$\Delta I_{SH}(\alpha) = 4\, I_{SH,0}(0)\, \gamma_0\, \frac{\Lambda(\delta_{ph}, L)}{L} \cos^3 \alpha \sin \alpha, \tag{3}$$

where $I_{SH,0}(0)$ is the equilibrium SH intensity at $\alpha = 0$, and $\gamma_0 = \kappa(\theta)\, r_{22} Q_0$ is the maximum rotation angle of the optical axes induced by the phonon amplitude $Q_0$ at the fixed delay. The 4-fold dependence on the input polarization, $\sim \cos^3 \alpha \sin \alpha$, is consistent with the alternating sign and angular symmetry observed experimentally, and the strength of the effect depends on the BBO thickness, $L$ and $\Lambda(\delta_{ph}, L)$, which includes the phonon penetration depth.

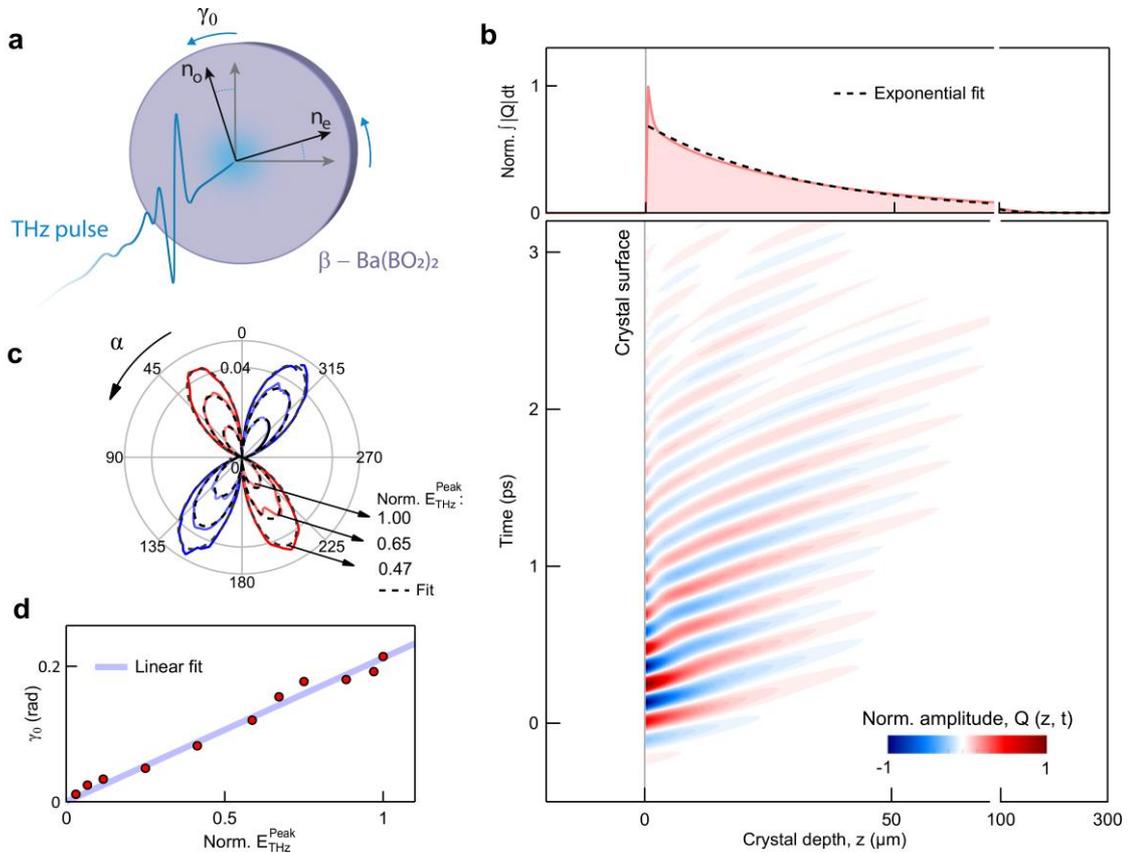

*Fig. 4. Theoretical modeling of SHG modulation and FDTD simulation for the phonon propagation. **a.** The BBO principal axes with corresponding refractive indices $n_o$ and $n_e$ undergo a rotation in the laboratory frame by an angle $\gamma_0$ due to the lattice displacement Q. **b.** Spatio-temporal evolution of the normalized (norm.) THz-driven phonon amplitude $Q(z,t)$ inside the bulk BBO crystal calculated using FDTD method. Top panel: FDTD results for the time-integrated phonon amplitude Q as a function of depth in the crystal (red curve) and the exponential fit of the decay, $Q(z) = Q_0 e^{-z/\delta_{ph}}$ (black dotted curve), to estimate the phonon penetration depth $\delta_{ph}$. **c.** The phonon-driven rotation of the principal axes, $\gamma_0$, results in a polarization-dependent SHG modulation with $\cos^3(\alpha)\sin(\alpha)$ angular dependence (Eq. 2). Dashed black lines are fits (based on Eq. 3) of the measurements as a function of the normalized THz field*

*strength at a fixed delay of 2.4 ps between NIR probe and THz pump-pulse.* ***d.*** *Estimate of $\gamma_0$ as a function of peak THz pulse field strength using the fits to polarization resolved SHG modulation measurements.*

A key parameter is the phonon penetration depth $\delta_{ph}$, which can be estimated independently using finite-difference time-domain (FDTD) methods to calculate the spatio-temporal dynamics of the phonon amplitude. With FTIR measurements of the linear optical properties of the BBO crystal, EOS measurements of the THz driving electric field, and the phonon equation of motion, the normalized amplitude $Q(z,t)$ of the phonon at $\omega_Q$ is simulated using FDTD and shown as a function of the crystal depth and time in Fig. 4b. Integrating the phonon amplitude in time and fitting the depth-dependent profile with an exponential decay (Fig. 4c) yields a phonon penetration depth of $\delta_{ph}$ =30.2 μm.

Using $\delta_{ph}$, angularly dependent SHG measurements (at fixed time delay) for THz pulses of varying intensity are then fit with Eq. 3, keeping $I_{SH,0}(0)$ as a constant parameter. These fits yield $\gamma_0$ as a function of the THz field amplitude (Fig. 4d-e). The rotation of the ordinary and extraordinary axes, $\gamma_0$, is found to vary linearly with applied THz field amplitude, consistent with the linear dependence of the IR-active phonon amplitude $Q_0$ on the THz field strength. The value of $\gamma_0$ achieved for the maximum THz field strength in the experiment is $\gamma_{0,max} = 0.216\ rad$.

**Microscopic Mechanisms of Phonon-Driven Nonlinear Optical Modulation**
The polarization-resolved measurements in conjunction with the analytical analysis yield the rotation of the optical axes due to the resonant THz-phonon excitation. This, in turn, permits quantification of the associated modulation of the refractive indices $\delta n$ in the crystal coordinate system, but does not immediately allow for independent retrieval of the electro-optic coefficient $r_{22}^E$ or a determination of the phonon displacement $Q$. At the microscopic level, the effect of the THz-induced coherent lattice vibrations on the local electronic environment within the anionic units remains hidden, as does the electro-optic coefficient $r_{22}^E$.
To access this information density functional theory (DFPT) can be used in the frozen-phonon approximation, to calculate the electronic band structure[32] - and, by extension, the dielectric function corresponding to a lattice displacement $Q_0$ along the phonon coordinates. The resulting dielectric function can then be used to extract absolute values for the refractive indices, $n_{x'}^{THz}$ and $n_{y'}^{THz}$ in the $(x',y',z')$ coordinate system of the crystal as a function of the phonon displacement $Q_0$ (Fig. 5a). Starting from equilibrium ($Q_0 = 0$), both indices are found to be $n_{x'} = n_{y'} = 1.705$, in good agreement with experimental measurements and previous ab-initio calculations[32]. With a finite phonon amplitude $Q_0$, the refractive indices respond linearly as expected: $n_{x'}^{THz}$ increases, while $n_{y'}^{THz}$ decreases with increasing $Q_0$ (see Fig. 5a).

Using the DFPT-calculated dependence of $\delta n$ on $Q_0$, the phonon-optic coupling coefficient $r_{22}$ can be estimated from the relation, $r_{22} = 2\delta n(Q_0)/n_{x'}^3 Q_0$, yielding a value of $r_{22} = 0.0049$ Å$^{-1}$(amu)$^{-1/2}$. This coefficient is especially relevant, as it can be used to calculate the maximum phonon amplitude at maximum THz field strength using the experimentally determined rotation of the optical axes $\gamma_{0,max}$ at the crystal surface, $Q_{0,max} = \gamma_{0,max}/(\kappa\, r_{22}) = 0.67$ Å(amu)$^{1/2}$ (Fig.5a). This value is consistent with expectations based on experiments targeting different effects in various materials that are nevertheless similar in that they also require THz excitation of large amplitude phonons. In these independent experiments, performed at X-ray free-electron lasers, it was possible to make a direct structural determination of the phonon amplitude[35,43]. Finally, the electro-optic coefficient $r_{22}^E(\omega)$ can be deduced from phonon-optic coupling coefficient $r_{22}$ using the conventional linear relation between the THz field and $Q$ in the frequency domain, which leads to the following expression:

$$r_{22}^E(\omega) = r_{22}^E(0) \frac{(\omega_Q^2 - \omega^2)}{(\omega_Q^2 - \omega^2)^2 - (\Gamma\omega)^2}, \qquad (4)$$

where $\omega_Q$ and $\Gamma$ are the phonon linewidth and the $r_{22}^E(0)$ is the DC electro-optic coefficient (see Sec. 9 of the Supplementary Information for details). Approaching the phonon resonance, the $r_{22}^E(\omega)$ peak value is 55 $pm/V$ at $\omega_* = \sqrt{\omega_Q^2 - \Gamma\omega_Q} = 4.22$ THz. Notably, the spectrum of the time-dependent SHG modulation dynamics closely matches the EO coefficint, with shared maxima just below $\omega_Q$ (dashed line Fig. 5b). This frequency dependence provides further evidence that the SHG modulation arises from the electro-optic effect. Indeed, in phonon-mediated EO processes, the response is governed not by the imaginary part of the dielectric function, $\varepsilon_2(\omega)$, which peaks at the phonon frequency $\omega_Q$, but by the real part, $\varepsilon_1(\omega)$, whose maximum occurs at $\omega_*$[44].

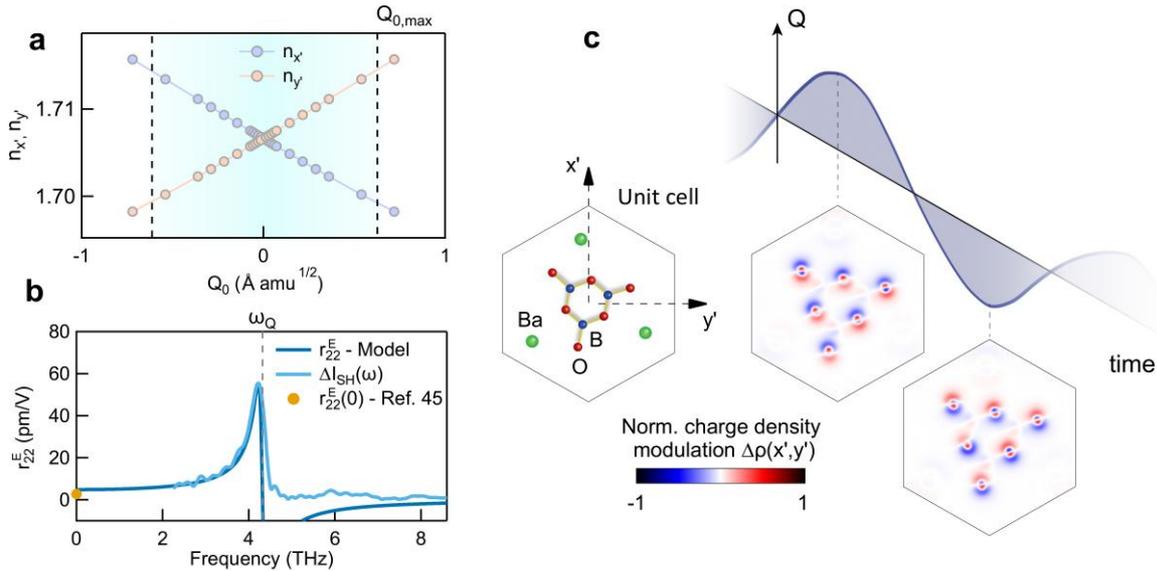

*Fig. 5. Frozen-phonon DFPT Calculation Results and Phononic-modulation Spatial Charge-density modulation. a. Computed refractive indexes $n_{x'}^{THz}$ and $n_{y'}^{THz}$ along the x' and y' axes of the unit cell as a function of $Q_0$. Vertical dashed*

lines, which delimit the shaded light-blue region, indicate the maximum amplitude estimated in the experiment, $Q_{0,max} \sim 0.67$ Å(amu)$^{1/2}$ at 2.4 ps. **b.** Frequency dependenc of the electro optic coefficient $r_{22}^E(\omega)$, shown together with the spectrum of the SHG-modulation dynamics for comparison. **c.** Planar modulation of the normalized (norm.) charge density $\Delta\rho(x', y')$ in the unit cell of β-BaB$_2$O$_4$ induced by phonon displacement for $Q_{0,max}$ and $-Q_{0,max}$. The electronic density is significantly affected by the O atomic displacement of the B$_3$O$_6$ unit.

The peak value of the vibrational electro-optic coefficient can also be estimated experimentally by combining the analytical expression for the SHG modulation with the phonon equation of motion (see Sec. 9 of the Supplementary Information). Using this approach and the broadband THz waveform applied in the experiment, a peak value of $r_{22}^E$ of $\sim 140\ pm/V$ is obtained in relatively good agreement with that obtained by DFPT (see Supplementary Fig. 7). To exclude the involvement of other phonon modes, additional experiments were performed using a narrowband THz pulse, which yields a similar value of ~170 pm/V (see Supplementary Fig. 8). The discrepancy of less than a factor of 3 between theory and experiment, which exists in both the broadband and narrowband case could therefore stem from the calibration of the absolute THz field amplitude, phonon parameters, and on the optical constants involved in the analysis—including their large frequency dispersion near the phonon resonance and dependence on the geometrical parameters in the experiment.

On the other hand, the theoretical treatment could also be slightly flawed, as extrapolating the theoretical electro-optic coefficient to the DC limit gives $r_{22}^E(0) = 6.6\ pm/V$, which is larger than the commonly reported experimental DC value of approximately 2.5 pm/V for BBO [45].

Regardless of the small discrepancy that remains, relative agreement between the experimental and theoretically determined electro-optic coefficient lends credence to the DFPT calculations, by which additional parameters can be obtained in the vicinity of the phonon resonance. In particular, the interaction between the phonon and the local electronic environment can be explored computationally. Here, DFPT-calculated differential charge density maps $\Delta\rho(x', y')$ confirm that the phonon mode predominantly drives cooperative displacements of the anionic [B$_3$O$_6$]$^{3-}$ rings within the lattice plane, while the Ba$^{2+}$ cations remain largely stationary (Fig 5c). This selective atomic motion produces a marked redistribution of charge density around the planar [B$_3$O$_6$]$^{3-}$ units. The electronic density around the oxygen atoms is largely influenced by the phonon dynamics, suggesting that the local electronic environment of these atoms plays a central role in modulating the refractive indices $n_{x'}^{THz}$ and $n_{y'}^{THz}$.

**Discussion**

Coherent lattice dynamics driven by intense THz pulses tuned to resonance with an infrared-active phonon mode have been shown to strongly modulate SHG in BBO on ultrafast timescales with the phase of the driving optical field. Analytical modeling, and first-principles DFPT calculations,, provide both a macroscopic overview and a comprehensive description of the physics at the microscopic level.

At the atomic level, DFPT calculations indicate that the THz driven phonons induce acentric displacements of the $B_3O_6$ anionic rings in the BBO unit cell. These displacements are phase-locked to the phonon dynamics (and driving field) and estimated to be as large as ~ 0.5 $Å^{-1}$ $(amu)^{-1/2}$, leading to a redistribution of local charge around the oxygen atoms. The impact of the transient charge redistribution on the refractive index ellipsoid of BBO is quantified through DFPT calculations. Additionally, the phonon-mediated electro-optic rotation of the ordinary and extraordinary optical axes that governs SHG is quantified using an analytical model. The observed SHG-modulation spectrum and angular dependence are consistent with the predicted EO response, exhibiting a resonance immediately below the IR-active phonon frequency. From a macroscopic perspective, it is the link between the rotation of the optical axes – which are defined at equilibrium by the experiment geometry – and the optical phase-matching conditions that permit strong modulation of the SHG signal in the bulk material that exceeds 30%.

Excellent agreement between theoretical predictions and experimental observations show that an approach that combines microscopic computational and macroscopic analytic treatment can be leveraged to capture the interplay between phonons and electronic states underlying nonlinear optical phenomena in solids[30]. These findings also establish DFPT as a powerful framework in which the optical properties of out-of-equilibrium crystal structures can be accurately evaluated.

Beyond fundamental insight, these findings may impact the design of new materials with tailored transient nonlinear responses, advancing ultrafast photonics and THz technologies. The ability to dynamically modulate SHG with high efficiency and precision through lattice-driven mechanisms opens new opportunities for ultrafast optical signal processing, high-speed data communication, and nonlinear optical switching. By leveraging lattice degrees of freedom[46,47], this work introduces a resonant pathway for ultrafast, high-efficiency control of nonlinear optical responses in three-dimensional bulk materials. Further, by general application of DFPT as a predictive tool for the optical-electronic-lattice interaction, combined with analytical treatment and experiments in the future, it can be expected that a broad range of nonlinear optical responses in other materials may be optimised and tailored for specific functionalities.

## Methods
### Experimental details
Intense THz pulses were generated through optical rectification in a DSTMS organic crystal using the signal beam of an optical parametric amplifier, driven by 50 fs, 800 nm pulses at a 100 Hz repetition rate. The SHG dynamics in β-$BaB_2O_4$ were investigated using a time-delayed replica of the 800 nm pulses. The experimental setup is shown in Supplementary Fig. 1; further details are provided in Sec. 1 of the Supplementary Information.

All experiments were performed at room temperature. The sample used was a commercially available β-BaB₂O₄ single crystal (5 mm × 5 mm × 0.3 mm), oriented at $\theta = 29.3°$ and $\phi = 90°$. Crystal orientation and crystallographic quality were verified by Laue X-ray diffraction (Supplementary Fig. 4).

**FTDT Spatio-Temporal Simulations of Phonon Propagation**

The THz-driven phonon propagation along the crystal depth has been simulated in time domain by computing Maxwell's equations based on FTDT method[48]. The material optical response along the o-axis at THz frequencies was modeled including the dielectric function of IR-active phonons, whose parameters have been experimentally determined using FTIR spectroscopy (see Sec. 3 of the Supplementary Information). The dominant feature is the phonon at $\omega_Q$, but we include additional 13 phonons for a faithful description of the optical response (phonon parameters are listed in Supplementary Table 1). For each $i$-th phonon mode, the equation of motion in the time domain is given by[35]

$$\ddot{Q}_i + 2\Gamma_i \dot{Q}_i + \omega_{Q,i}^2 Q_i = Z\, E_{THz}(t) \tag{5}$$

Where $\Gamma_i$ is the damping rate, $\omega_{Q,i}$ is the phonon frequency and $Z = \omega_{Q,i}\sqrt{\epsilon_0(\varepsilon_0 - \varepsilon_\infty)}$ is the phonon effective charge, being $\varepsilon_0$ and $\varepsilon$ the dielectric constant at the low and high-frequency limit respectively and $\epsilon_0$ the vacuum permittivity.

Lattice equations and the electric field component from Maxwell's equations in media are related by displacement field as:

$$D = \epsilon_0 \varepsilon_\infty E_{THz} + \sum_i \omega_{Q,i}\sqrt{\epsilon_0(\varepsilon_0 - \varepsilon_\infty)}\, Q_i \tag{6}$$

By FTDT calculation we numerically solve the above equation at points of the grid through leapfrog time-step scheme. In the simulation, the THz electric field $E_{THz}(t) = E_0 \cos(\omega_0 t)\, e^{-\frac{1}{2}\frac{t^2}{\sigma^2}}$ is a Gaussian field having a width $\sigma = 100$ fs and a center frequency $\omega_0 = 5$ THz. The plot of the spatiotemporal propagation of the phonon at $\omega_Q$ is show in Fig. 4b.

**DFPT Calculations of Linear and Nonlinear Optical Properties Under Phonon Displacement**

To quantify the linear and nonlinear optical responses due to phonon displacement, first-principles calculations were perfomed in the framework of density-functional perturbation theory (DFPT) implemented using PHonon package[49] in Quantum ESPRESSO[50,51], with PBE SG15 Optimized Norm-Conserving Vanderbilt (ONCV) pseudopotentials[52,53].

DFPT calculations were conducted within the structural unit cell [42 atoms, R3c (No. 161) space group] of the BBO crystal at room temperature has been taken from Ref.[54]. For the Brillouin zone sampling, 4x4x4 k-point grid and a plane-wave energy cutoff of 80 Ry have been used and the structural parameters have been relaxed until the force acting on each atom was less than 0.0025 eV/Å and the pressure was below 0.5 kbar. Though phonon energies and corresponding displacement vectors at the Brillouin zone center, the macroscopic dielectric tensor, and Born effective charges were computed.

The refractive index ellipsoid were computed as the square-root of the macroscopic dielectric tensor, calculated with DFPT in the long-wavelength limit ($q = 0$). This is valid in non-absorbing materials, in which the refractive index dispersion is small, as in the case of BBO in the optical optical region.

At zero displacement along the phonon coordinates ($Q_0 = 0$), the DFPT-calculated dielectric tensor in the crystal frame is:

$$\xi_0 = \begin{pmatrix} \xi^0_{x'x'} & 0 & 0 \\ 0 & \xi^0_{x'x'} & 0 \\ 0 & 0 & \xi^0_{z'z'} \end{pmatrix}. \tag{7}$$

which corresponds to a refractive $n_{x'} = \sqrt{\xi_{x'x'}} = 1.705$ and $n_{z'} = \sqrt{\xi_{z'z'}} = 1.705$. Previous DFPT studies have reported a refractive index of $n_{x'} = n_{y'} = 1.692$ at 800 nm, see Ref.[32], which is within 0.5% of our DFPT-calculated value. This close agreement validates the use of the long-wavelength approximation in our approach. Additionally, the small discrepancy might by due to the fact our DFPT method, Ref.[49], includes local-field effects and goes beyond the independent-particle approximation employed in Ref. [32]. Experimentally, the refractive index has been measured to be $n_{x'} = 1.66$, which is within 3% of our calculated results.

The dielectric tensor as a function of the phonon amplitude, was computed with the same approach but assuming structures displaced along the relevant IR-active phonon mode displacement vectors.

The dielectric tensor as a function of the phonon amplitude has the following form:

$$\xi_Q = \begin{pmatrix} \xi_{x'x'} & \xi_{x'y'} & 0 \\ \xi_{x'y'} & \xi_{x'x'} & 0 \\ 0 & 0 & \xi_{zz} \end{pmatrix} \xrightarrow{R(\frac{\pi}{2})} \begin{pmatrix} \tilde{\xi}_{x'x'} & 0 & 0 \\ 0 & \tilde{\xi}_{x'x'} & 0 \\ 0 & 0 & \xi_{z'z'} \end{pmatrix} \tag{8}$$

which diagonalized with a $x' - y'$ plane $\frac{\pi}{2}$ rotation around the $z'$ axis, as in the analytical model (see Sec. 6 of the Supplementary Information). For the dielectric tensor in Eq. 7, the refractive indeces as a fuction of phonon amplitude were calculated.

For the electronic density calculations in Fig. 4b, a plane-wave kinetic energy cutoff of 320 Ry has been used. To evaluate the nonlinear optical coefficients of BBO (see Supplementary Fig. 6) finite-field approach was used[55]. In this method, the electric field is explicitly included in the DFPT calculations, and the nonlinear coefficients are obtained numerically by computing the second derivative of the polarization with respect to the applied electric field.

**Acknowledgments.** The authors thank the PSI-LNO GL group for support during the experiments and R. Li for helpful discussions. This work was supported by the Swiss National Science Foundation (SNSF) under grant number 10001644 and SNSF Spark grant number 221173.

**Author Contributions Statement.** F.G. and C.V. conceived the project and designed the experiment. F.G., C.V. and G.M. performed the THz spectroscopy measurements. F.G., C.V., A.T., G.N. and L.S. carried out the THz experiments. F.G. performed the FDTD simulations. N.C. carried out the DFPT calculations. N.F., N.C. and F.G. developed the analytical model and interpreted the data. F.G. prepared the figures. F.G., A.L.C. and C.V. wrote the manuscript with input from all authors. All authors discussed the results.

**Competing Interests Statement.** The authors declare no competing interests.

# Supplementary Information: All-optical control of second-harmonic generation in β-BaB₂O₄ via coherent, terahertz-driven acentric lattice displacement

Flavio Giorgianni[1,2,*], Nicola Colonna[2], Gabriel Nagamine[1], Leonie Spitz[2], Guy Matmon[2], Alexandre Trisorio[2], Nicolas Forget[3], Carlo Vicario[2], Adrian L. Cavalieri[1,2]

[1]*Institute of Applied Physics, University of Bern, CH-3012 Bern, Switzerland*
[2]*Paul Scherrer Institute, CH-5232 Villigen-PSI, Switzerland*
[3]*Institut de Physique de Nice (INPHYNI), Université Côte d'Azur*

## 1. Experimental Setup and THz Field Characterization

A schematic of the experimental geometry is shown in Supplementary Fig. 1a. The measurements are made using a 100 Hz, 20 mJ, 50 fs Ti:Sa laser system coupled to an optical parametric amplifier (OPA). The output energy of the OPA signal beam was approx. 3.9 mJ per pulse. The OPA signal, at a wavelength of 1.5 μm, was employed to generate intense single-cycle THz pulses through optical rectification in a large-aperture DAST crystal, with clear aperture exceeding 10 mm in diameter. The OPA signal beam diameter was approximately 9 mm at the crystal surface. Metallic multi-mesh filters were inserted into the THz optical path to isolate the THz pulses from the residual OPA beam and to shape the THz pulse spectrum.

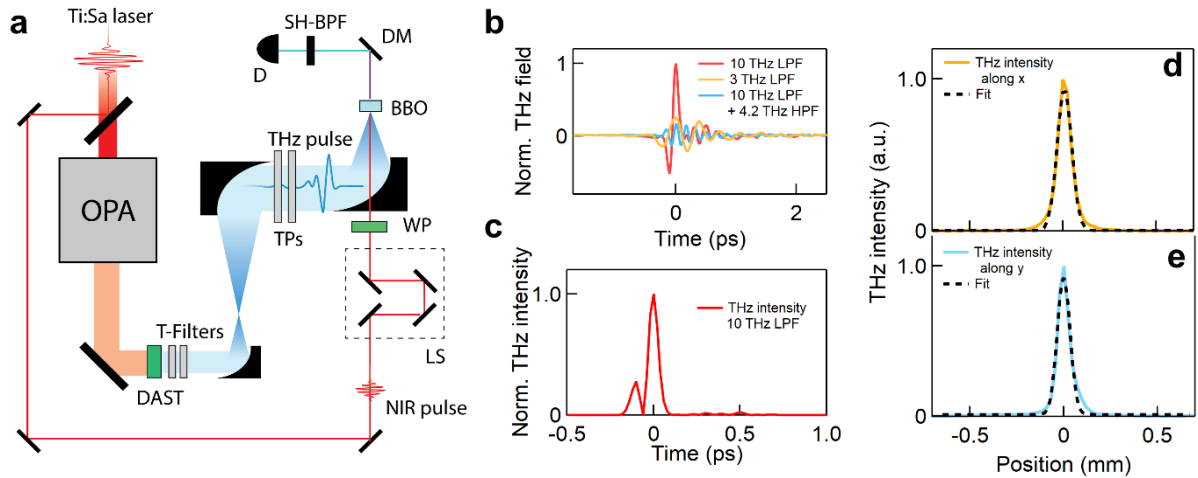

*Supplementary Fig. 2. Experimental Setup and THz field characterization.* **a**, *Setup for THz generation and modulation of second harmonic in BBO. LS: linear stage; WP: half-wave plate, DM, dichroic mirror, SH-BPF: band-pass filter for SH signal, T-Filter: THz spectral filters, TPs: THz polarizers, BBO: BBO crystal sample*. **b**, *THz waveforms for 10 THz LPF, 3 THz LPF, and 4.2 THz HPF+10 THz LPF.* **c**, *Intensity profile $|E(t)|^2$ of the THz pulse with the 10 THz LPF.* **d**, *Horizontal and* **e**, *vertical THz beam transverse profile of THz pulse at the sample position with the 10 THz LPF. Orange and light blue solid curves are the experimental data measured with a THz camera while black dotted lines represent a Gaussian fit. The estimated FWHM are 90 μm and 85 μm, respectively.*

The THz beam was expanded and then focused onto the β-BaB₂O₄ (BBO) crystal sample using three off-axis parabolic mirrors with respective focal lengths of 1", 6", and 2" (see Supplementary Fig. 1a), which permits tight focusing of the THz beam onto the sample. A pair of wire-grid THz polarizers are used to vary the THz electric field strength at the sample (TPs in Supplementary Fig. 1a). The rotation angle of first polarizer is adjustable, while the second polarizer was fixed to transmit only the vertically polarized component.
For the broadband THz-pump experiments (see Fig. 1c–f in the main text), isolation of the OPA signal beam was achieved using a filter set consisting of two 20 THz low-pass filters (LPFs) and one 10 THz LPF (all from QMC Ltd). This configuration provided a measured extinction ratio of $\sim 10^7$ for the residual 1.5 μm OPA pulse. To selectively tailor the THz pump spectrum relative to the phonon resonance (see Fig. 2a of the main manuscript), the 10 THz LPF in the filter set was replaced as follows: (i) for isolating frequency components below the phonon resonance, a 3 THz LPF was used; (ii) for isolating components above the resonance, a combination of a 4.2 THz high-pass filter (HPF) followed by a 10 THz LPF was used. The temporal profile of the THz pulse at the sample position was measured using electro-optic sampling (EOS) with a (110) GaP crystal of 200 μm thickness. The THz waveforms measured by EOS for these three filter configurations are shown in Supplementary Fig. 1b.

The broadband THz electric field strength was estimated from measurement of the temporal profile of the THz pump, the THz pulse energy, and the focused THz spot size. The intensity temporal profile of the broadband pump pulse computed as the square of the THz electric field waveform is shown in Supplementary Fig. 1c. The transverse spatial profile of the THz beam was measured using a THz micro-bolometric camera (NEC IRVT0830). The THz energy per pulse was measured with a Gentec THZ12D-3S-VP-D0 energy meter.

The THz pulse in Fig. 1 of the main manuscript has the following parameters: THz energy/pulse: 3.5 $\mu J$, focused spotsize (FWHM): 90 $\mu m$ from Gaussian fit of the beam profile (see Supplementary Fig. 1d-e).

A THz field strength, $E_{THz} \approx 8.5\ MV/cm$, was estimated using the following expression[1]:

$$E_{THz} = \sqrt{\frac{z_0 W}{\pi w^2 \int_{-\infty}^{\infty} g^2(t)dt}} \qquad (S1)$$

where $z_0 = 377\ \Omega$ is the free space impedance, $W$ is the THz energy/pulse, $w$ is the beam waist and $g^2(t)$ is the temporal intensity profile of the THz pulse with peak value normalized to 1, (see Supplementary Fig. 1b). The temporal profile $g(t)$ was measured by EOS and the integral, $\int_{-\infty}^{\infty} g^2(t)dt = 0.086$ ps. The THz field strengths for pump spectra above and below the phonon resonance were determined from EOS measurements. The THz pulse below the phonon resonance (obtained using the 3THz LPF), has a peak electric field that is 4 times smaller than the peak electric field of the broadband THz pulse obtained with the 10 THz LPF alone (see Supplementary Fig. 1b). Above resonance (implemented using the 4.2 THz HPF in combination with the 10 THz LPF), the peak THz field amplitude was 0.16 times the peak field of the broadband THz pump pulse.

## 2. Evidence of THz-field-induced second harmonic intensity modulation in BBO

To verify that the observed SH intensity modulation does not result from variations in the intensity of the transmitted fundamental frequency (FF) beam potentially due to THz-induced changes in the Fresnel coefficients, the transmitted NIR pulse intensity was measured as a function of the time delay relative to the THz pump pulse (see Supplementary Fig. 2). No significant modulation is observed in the NIR pulse intensity, confirming that the SHG modulation originates from the resonant THz-driven phonon excitation.

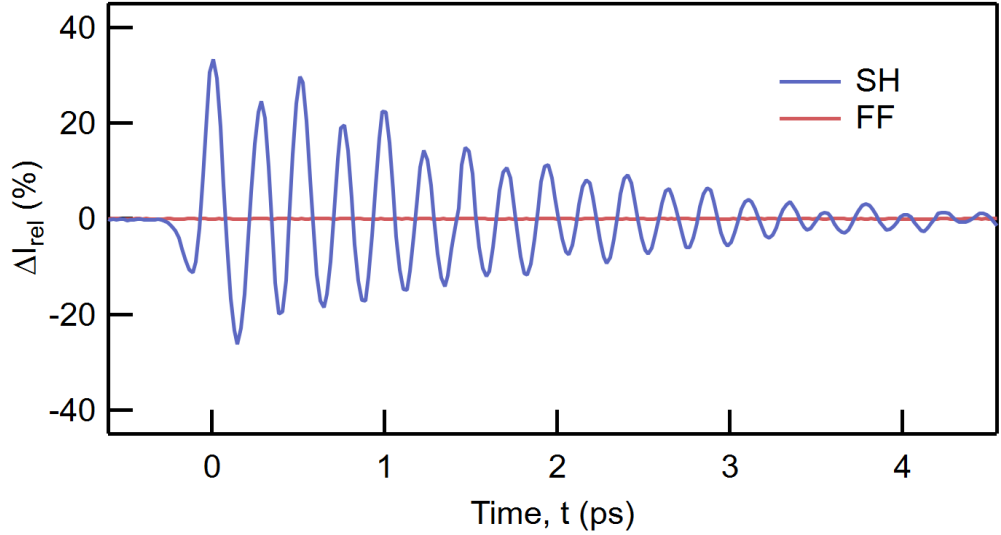

*Supplementary Fig. 2. Experimental measurement of THz-field-induced modulation of SH and FF intensities. The absence of significant modulation in the FF beam intensity confirms that the observed modulation effect originates from the SH process.*

## 3. THz Spectroscopy of BBO and Laue X-ray Diffraction

To characterize the BBO infrared-active phonons, optical transmission measurements in the THz regime were performed using Fourier transform infrared (FTIR) spectroscopy at the IR beam line of the Swiss Light Source at PSI. For these measurements, a 30-μm-thick reference BBO crystal was used with the same orientation ($\theta = 29.3°$ and $\phi = 90°$) as the crystal in the main THz experiment, but with reduced thickness to maximize the THz transmission. Supplementary Fig. 3 shows the transmission spectra in the THz range for light polarised along the ordinary-axis, which is consistent with previously reported work[2–4].

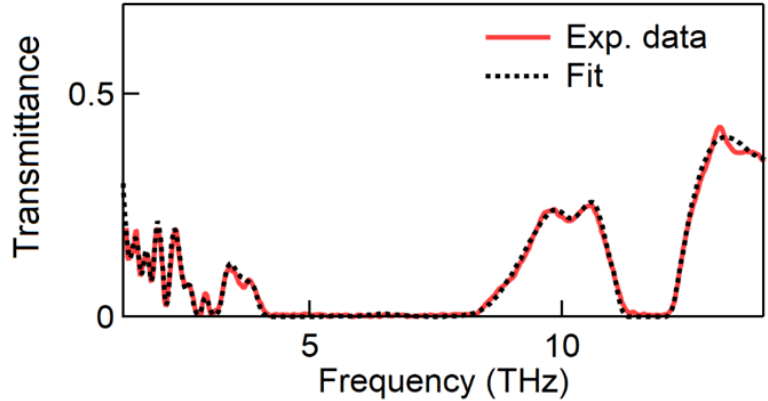

*Supplementary Fig. 3. Transmission spectrum in the THz regime of a 30 μm-thick BBO crystal measured by FTIR spectroscopy.* Black dotted line denotes the fit curve.

The real $\varepsilon_1(\omega)$ and imaginary part $\varepsilon_2(\omega)$ of complex dielectric function, depicted in Fig. 1d of the main manuscript, were found by fitting the optical spectra using the Kramers-Kronig (KK) consistent dielectric function model with Reffit software[5]. The fit procedure for the transmission spectrum was based on the conventional dielectric function:

$$\varepsilon(\omega) = \varepsilon_\infty + \sum_i \frac{\omega_{p,i}^2}{\omega_{o,i}^2 - \omega^2 - i\Gamma_i\omega} \tag{S2}$$

Where $\omega_{0,i}$ (cm⁻¹), $\omega_{p,i}$ (cm⁻¹) and $\Gamma_i$ (cm⁻¹) are the phonon strength (or plasma frequency), the eigenfrequency and the linewidth of the $i$-th Lorentz oscillator, respectively. $\varepsilon_\infty$ is the high frequency dielectric constant. The black dashed line in Supplementary Fig. 3 is the fit and the corresponding parameters are listed in Supplementary Table 1. These parameters were used to model the BBO optical response in the finite-difference time-domain (FDTD) simulations (see Eq. 5 in Methods).

| $i$ | $\omega_{0,i}$ (cm⁻¹) | $\omega_{p,i}$ (cm⁻¹) | $\Gamma_i$ (cm⁻¹) | $i$ | $\omega_{0,i}$ (cm⁻¹) | $\omega_{p,i}$ (cm⁻¹) | $\Gamma_i$ (cm⁻¹) |
|---|---|---|---|---|---|---|---|
| 1 | 47.14 | 44.33 | 8.63 | 8 | 119.62 | 41.72 | 14.68 |
| 2 | 55.29 | 33.98 | 5.82 | 9 | 144.85 | 217.02 | 6.49 |
| 3 | 61.91 | 33.59 | 5.17 | 10 | 188.78 | 102.10 | 34.17 |
| 4 | 71.80 | 41.77 | 4.34 | 11 | 245.15 | 176.46 | 60.90 |
| 5 | 84.077 | 36.30 | 6.14 | 12 | 340.26 | 21.03 | 19.23 |
| 6 | 92.26 | 60.61 | 5.82 | 13 | 383.20 | 185.15 | 5.33 |
| 7 | 103.29 | 67.29 | 7.25 | 14 | 469.30 | 45.65 | 47.70 |

$\varepsilon_\infty = 2.89$

***Supplementary Table 1. Fit parameters of $\varepsilon(\omega)$ (Eq. S2).*** The parameters of the phonon at $\omega_{Q1} = 4.32\ THz$ are highlighted in blue.

Phonon resonances are found at $\omega_{Q1} = 4.32$ THz (144 cm⁻¹); $\omega_{Q2} = 5.65$ THz (189 cm⁻¹), $\omega_{Q3} = 7.34$ THz (245 cm⁻¹) consistent with those in Ref.[4], which show peaks at $\omega_{Q1} = 4.65$ THz, $\omega_{Q2} = 5.65$ THz, $\omega_{Q3} = 7.34$ THz. For $\omega_{Q1}$, the discrepancy of 0.3 THz between the measurements here and those reported in Ref.[4] could originate from sample variation and/or instrument resolution. The current measurements use a 30 µm-thick uncoated BBO and the FTIR was performed with a resolution of 0.014 THz. While the measurements in Ref.[4] were performed using a BBO crystal with a 400 nm antireflection coating (which could also affect the THz transmission) and transmission spectra were collected with a lower frequency resolution (0.05 THz). The 30µm reference BBO in the current measurements was also evaluated using Raman spectroscopy, which yields $E$-symmetry modes at 143 cm⁻¹, 187 cm⁻¹ and 245 cm⁻¹, which closely matches (to within 2 cm⁻¹) the values reported in Ref.[6]. Sample quality and orientation of the 300µm thick BBO used in the experiment were verified by Laue X-ray diffraction. Establishing the crystal orientation is essential for defining the crystallographic coordinate system $(x', y', z')$ and its relation to the laboratory frame $(x, y, z)$, which is required for both the DFPT calculations based on BBO crystal structure and the analytical model. The Laue pattern shows that the crystallographic *a*-axis corresponds to the y of the laboratory frame (ordinary axis), which coincides with the $x'$-axis in the crystal frame since $\phi = 90°$(Supplementary Fig. 4a). The crystal is *R3c* symmetric, so the $E$ phonons are doubly degenerate. DFT calculations indicate that the THz-driven $E$ mode at 144 cm⁻¹ has two orthogonal in-plane dipole moments: $E_1$ along $x'$ and $E_2$ along $y'$ (Supplementary Fig. 4b). When the THz pump is polarised along the a-axis, the field couples only to the $E_1$ branch and the orthogonal $E_2$ branch is not excited.

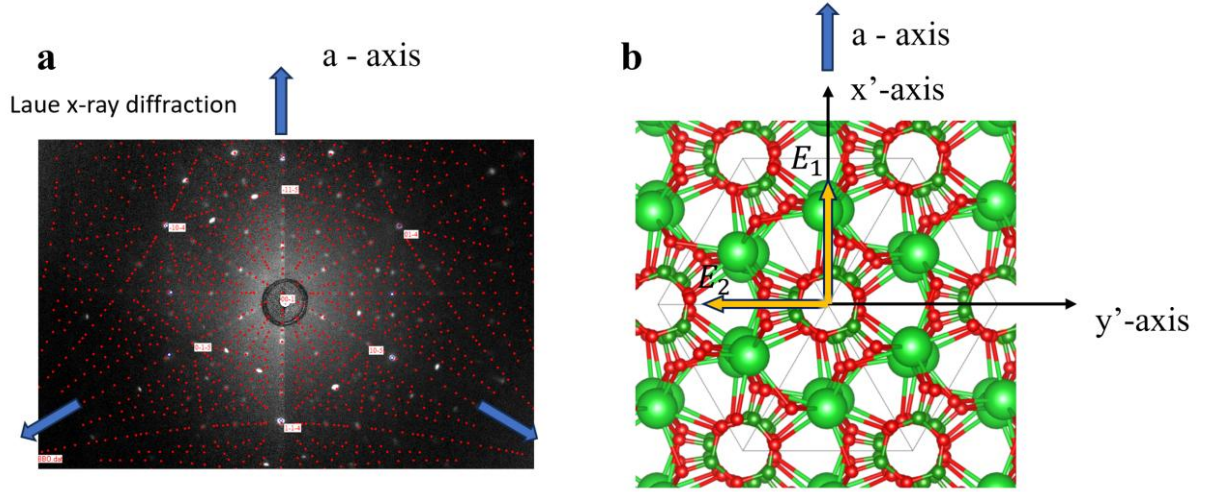

*Supplementary Fig. 4. Laue diffraction and sample orientation: a*, Laue diffraction pattern of the single crystal 300 μm thick BBO sample ($\theta = 29.3°$, $\phi = 90°$) used in the experiment to verify the experimental geometry. The THz electric field is aligned along the a-axis. In the Laue measurement, the crystal was rotated by ~ $-\theta°$ around the a-axis to confirm the $R3c$ symmetry. Red dots indicate simulated Laue diffraction peaks generated using Laue diffraction software. *b*, Crystal unit cell from DFPT calculations, shown in crystallografic coordinate system x', y' (in-plane) and z' (out-of-plane). The x'-axis corresponds to the crystallographic a-axis. The E-mode at 144 cm$^{-1}$ is doubly degenerate, with branches $E_1$ and $E_2$ having dipole moment directions indicated by the orange arrows.

## 4. Polarization angular dependence of static second harmonic generation

For BBO, the second-order nonlinear $d_{NL}$ tensor has 3 independent coefficients[7] that are defined in the crystal coordinate system $(x', y', z')$:

$$d_{NL} = \begin{pmatrix} 0 & 0 & 0 & 0 & d_{15} & -d_{22} \\ -d_{22} & d_{22} & 0 & d_{15} & 0 & 0 \\ d_{15} & d_{15} & d_{33} & 0 & 0 & 0 \end{pmatrix}, \quad \text{(S3)}$$

where $d_{15} = 0.16$ pm/V, $d_{22} = -2.2$ pm/V, $d_{33} \sim 0$.

In the laboratory coordinate system, the FF is linearly polarized and propagates in the z-direction. The polarization is defined by the angle $\alpha$ of the polarization vector with respect to the vertical, y-axis, which corresponds to the ordinary axis of the static BBO crystal in the laboratory frame. In this case:

$$E_{lab}^{FF}(\alpha) = E_0\, u, \quad \text{(S4)}$$

where $u = (\sin\alpha, \cos\alpha, 0)$ is the polarization vector and $E_0$ is the amplitude of the FF beam. Experimentally, the polarisation is varied using a half-wave plate.

The polarization vector in the crystal frame is found using the transformation $u_{xtal} = Ru$ where $\theta$ is the polar angle - which is the angle of the lab frame z-axis with respect to the crystal c-axis - and $\phi$ is the azimuthal angle -which is the angle of the lab frame x- and y-axes with respect to the crystal ordinary axis (for the BBO crystal used in the experiments, $\theta = 29.3°$ and $\phi = 90°$).

$$R = \begin{pmatrix} -\cos\theta\cos\phi & \sin\phi & \cos\phi\sin\theta \\ -\cos\theta\sin\phi & -\cos\phi & \sin\theta\sin\phi \\ \sin\theta & 0 & \cos\theta \end{pmatrix} \quad \text{(S5)}$$

With this transformation, it follows that the effective nonlinear tensor $d_{eff}$ in the laboratory frame, is:

$$d_{eff} = R^T d_{NL}\, f_{3\to6}(Ru), \quad \text{(S6)}$$

Where $f_{3\to 6}$ is the operator[8] that maps the square of the electric field components $(x', y', z')$ onto $(x'^2, y'^2, z'^2, 2y'z', 2x'z', 2x'y')$ in the crystal frame.

For Type-I phase matching ($o + o \to e$), the SH harmonic is polarized along the extraordinary axis ($x$-axis in the lab frame). Therefore, for this component one finds:

$$E_{SH}^x(\alpha) \propto (\sin\theta\, d_{15} - \cos\theta \sin 3\phi\, d_{22}) \cos^2\alpha\, E_0^2 \tag{S7}$$

For this particular SH component, the nonlinear tensor elements in $d$ can be written as a scalar coefficient $d_{eff}$.

$$E_{SH}^x(\alpha) \propto d_{eff} \cos^2\alpha\, E_0^2 \tag{S8}$$

The polarisation-dependent Type-I generated SH intensity $I_{SH,0}$ is then proportional to $d_{eff}^2$ and $\cos^4\alpha$ as experimentally observed (Fig. 3b in the main manuscript).

## 5. Analytical model: THz Vibrational Modulation of Optical Properties

The THz-driven phonon modulates the refractive index ellipsoid $\eta_i$ through the electro-optic tensor $r_{ij}$. The BBO crystal structure is *3m,* therefore the electro-optic tensor in the Voigt notation in the crystal coordinate system $(x', y', z')$ is written as[8]:

$$r_{ij}^{EO} = \begin{pmatrix} 0 & -r_{22} & r_{13} \\ 0 & r_{22} & r_{13} \\ 0 & 0 & r_{33} \\ 0 & r_{51} & 0 \\ r_{51} & 0 & 0 \\ -r_{22} & 0 & 0 \end{pmatrix}. \tag{S9}$$

With an applied field, the components of the index of refraction ellipsoid undergo a variation $\Delta\eta_i = r_{ij}^{EO} \cdot E_j$, where $E_j$ are the electric field components in the crystal frame driving the electro-optic effect. The electro-optic effect generated by a temporally modulated electric field can occur through two distinct mechanisms: electronic and vibrational that are described by "electronic" electro-optic coefficient $r_{ij}^{el}$ and the "vibrational" electro-optic coefficient $r_{ij}^{vib}$, Ref. [9]. When a THz electric field $E_{THz}$ is applied, the change in the index of refraction ellipsoid tensor elements can then be rewritten as: $\Delta\eta_i = (r_{ij}^{el} + r_{ij}^{vib}) \cdot E_{THz,j}$. In BBO, the vibrational contribution is known to be dominant – even for off-resonant electric fields at low frequencies and it is strongly enhanced when approaching the phonon frequency at ~ 4.3 THz.[9] Evidence of this effect is found in the experiment through the observation of a significant increase in SH modulation on resonance, when the THz pump spectral components overlap with the phonon frequency (2a-b of the main manuscript). Therefore, near the phonon resonance, the vibrational contribution can be considered to be dominant and only $r_{ij}^{EO} = r_{ij}^{vib}$ is considered in the electro-optic tensor[9]. Since the phonon amplitude scales as $Q_j \propto ZE_{THz,j}$, where $Z$ is the phonon effective charge, the vibrational term $r_{ij}^{vib} E_{THz,j}$ can be rewritten as $r_{ij} Q_j$. Here, $r_{ij}$ is the phonon–optic coupling coefficient, which describes the modulation of the refractive-index ellipsoid element as a function of the phonon amplitude.

In the laboratory frame $(x, y, z)$, the extraordinary axis is along $x$ and the ordinary axis is along $y$, while the light propagates along the $z$ direction. In the experiment, the THz electric field polarisation is fixed and aligned along the ordinary, y-axis. The THz polarization in the Jones formalism is $E_{THz} = (0, E_{THz,y}, 0)$. The transformation matrix $R$ is used to convert the THz components from the laboratory frame $(x, y, z)$ to the crystallographic frame $(x', y', z')$:

$$R = \begin{pmatrix} -\cos\theta\cos\phi & \sin\phi & \cos\phi\sin\theta \\ -\cos\theta\sin\phi & -\cos\phi & \sin\theta\sin\phi \\ \sin\theta & 0 & \cos\theta \end{pmatrix} \quad (S10)$$

The THz electric field in the crystal frame is then $E_{THz}^{xtal} = R\,E_{THz}$. Under coherent phonon excitation, the index ellipsoid becomes $\eta_Q = \eta_0 + \Delta\eta_Q$, where $\eta_0$ is the index ellipsoid at the equilibrium [7]. In crystal frame $(x', y', z')$, $\eta_0$ can then be written in matrix form as:

$$\eta_0 = \begin{pmatrix} \frac{1}{n_{x'}^2} & 0 & 0 \\ 0 & \frac{1}{n_{x'}^2} & 0 \\ 0 & 0 & \frac{1}{n_{z'}^2} \end{pmatrix}, (n_{x'} = n_{y'}). \quad (S11)$$

Given that BBO is uniaxial, for the modulation of the index ellipsoid $\Delta\eta_Q$, only the phonon-optic coupling coefficient $r_{22}$, which is related to $r_{ij}^{EO}$, is significant and the others can be neglected[8,10].

As the phonon dipole moment is along the THz electric field direction[3], the phonon amplitude is $Q_0 \propto Z\,E_{THz}^{xtal} \equiv Z E_{THz,y'}$. Thus, the modulated refractive index ellipsoid, calculated from the electro-optic tensor in Eq. S9, can be written in the matrix form as:

$$\eta_Q = \begin{pmatrix} \frac{1}{n_{x'}^2} + r_{22}Q_0\cos\phi & -r_{22}Q_0\sin\phi & 0 \\ -r_{22}Q_0\sin\phi & \frac{1}{n_{x'}^2} - r_{22}Q_0\cos\phi & 0 \\ 0 & 0 & \frac{1}{n_{z'}^2} \end{pmatrix}. \quad (S12)$$

Since $\phi = 90°$, a finite value of phonon amplitude $Q_0$ leads to the emergence of off-diagonal terms, and the crystal becomes biaxial associated with an optical anisotropy in the $x'$-$y'$ plane, as the degeneracy between $x'$ and $y'$ directions is lifted.

The index ellipsoid in the crystal frame can be written as $\eta_Q = \eta_0 - r_{22}Q_0 \cdot A$, where:

$$A = \begin{pmatrix} 0 & 1 & 0 \\ 1 & 0 & 0 \\ 0 & 0 & 0 \end{pmatrix}. \quad (S13)$$

The dynamical refractive index ellipsoid in the laboratory frame $\eta_{THz}^{Lab}$ can then be determined upon cartesian axis transformation: $\eta_{THz}^{Lab} = R^T \eta_0 R - r_{22}Q_0\,R^T A R = R^T \eta_0 R - \epsilon\, R^T A R$, with $\epsilon = r_{22}Q_0$ and $R^T$ the transpose of $R$. The two terms of $\eta_{THz}^{Lab}$ are analysed separately:

$$R^T\eta_0 R = \begin{pmatrix} \frac{\cos^2\theta}{n_{x'}^2} + \frac{\sin^2\theta}{n_{z'}^2} & 0 & \cos\theta\sin\theta\left(\frac{1}{n_{z'}^2} - \frac{1}{n_{x'}^2}\right) \\ 0 & \frac{1}{n_{x'}^2} & 0 \\ \cos\theta\sin\theta\left(\frac{1}{n_{z'}^2} - \frac{1}{n_{x'}^2}\right) & 0 & \frac{\sin^2\theta}{n_{x'}^2} + \frac{\cos^2\theta}{n_{z'}^2} \end{pmatrix}, \quad (S14)$$

and
$$\epsilon R^T A R = \epsilon \begin{pmatrix} 0 & -\cos\theta & 0 \\ -\cos\theta & 0 & \sin\theta \\ 0 & \sin\theta & 0 \end{pmatrix}. \tag{S15}$$

With the definition $\frac{1}{n_e(\theta)^2} = \frac{\cos^2\theta}{n_{x\prime}^2} + \frac{\sin^2\theta}{n_{z\prime}^2}$ and $\frac{1}{n_o^2} = \frac{1}{n_{x\prime}^2}$, and reducing the analysis to two dimensions, considering that the light polarization lies in the x-y plane of the laboratory frame, it follows that:

$$\eta_{THz}^{Lab} = \begin{pmatrix} \frac{1}{n_e(\theta)^2} & \epsilon\cos\theta \\ \epsilon\cos\theta & \frac{1}{n_o^2} \end{pmatrix} \tag{S16}$$

This dynamical index ellipsoid in the lab frame is diagonalized by a rotation of $\gamma_0$, and the new refractive indexes are:

$$n_o^{THz} = n_o + (r_{22}Q_0)^2 n_o^3 \gamma_0/2 \tag{S17}$$

and

$$n_e^{THz} = n_e(\theta) - (r_{22}Q_0)^2 n_e^3(\theta)\, \gamma_0/2, \tag{S18}$$

with $\gamma_0 = \kappa\,\epsilon + O[\epsilon]^2 = \kappa\, r_{22}Q_0$ with $\kappa = \frac{\cos\theta}{\frac{1}{n_e(\theta)^2} - \frac{1}{n_o^2}}$.

The primary result is that the dynamical principal axes of the crystal in the laboratory frame are no longer aligned with the $x$ and $y$ crystal axes upon phonon excitation with amplitude $Q_0$. Instead, they are rotated by an angle $\gamma_0$, which is proportional to $Q_0$. Importantly, the magnitude of the refractive indices are not significantly altered by $Q_0$, since the correction is to $3^{\text{rd}}$-order $(r_{22}Q_0)^2\gamma_0 \propto (r_{22}Q_0)^3$.

## 6. Analytical model: Vibrational Modulation of Second Harmonic Intensity

For the Type-I phase-matched BBO crystal ($\phi = 90°, \theta = 29.3°$), the SHG intensity is proportional to $|E_y^{FF}|^4$, where $E_y^{FF}$ is the projection of the amplitude of the 800 nm FF on the lab-frame y-axis, which is parallel to the unperturbed crystal-frame ordinary axis. The SH will be polarised along the crystal extraordinary axis, which is parallel to the lab-frame x-axis.

A linearly polarised FF propagating along the z-direction in the laboratory frame has field components $(x, y)$ given by the following expression:

$$E^{FF}(\alpha) = E_0^{FF}(\sin\alpha, \cos\alpha) \tag{S19}$$

where $\alpha$ is the angle between that y-axis and the input polarisation.

Due to the birefringence of the BBO crystal, for input fundamental fields that are not aligned with the principal axes, the polarisation will evolve with propagation distance $\zeta$ in the crystal. In the static case, at equilibrium, after travelling a distance $\zeta$, the evolution of the field components is determined using the linear phase retardation matrix $\Gamma(\zeta)$ such that:

$$E^{FF}(\zeta, \alpha) = \Gamma(\zeta) E^{FF}(\alpha) \quad \text{where} \quad \Gamma(\zeta) = \begin{pmatrix} e^{i2\pi n_e \zeta/\lambda} & 0 \\ 0 & e^{i2\pi n_o \zeta/\lambda} \end{pmatrix} \tag{S20}$$

In this case, and in the undepleted pump approximation, the Type-I ($o + o \to e$) SH intensity from a crystal length $L$ as a function of input polarisation $\alpha$ can be written as:

$$I_{SH,0}(L, \alpha) \propto \left| \int_0^L \left( E_y^{FF}(\zeta, \alpha) \right)^2 e^{-ik_{n_e,2\omega}\zeta} d\zeta \right|^2 \tag{S21}$$

And upon substitution:

$$I_{SH,0}(L,\alpha) \propto \left| \int_0^L (E_0^{FF} \cos\alpha)^2 e^{i4\pi n_{o,\omega}\zeta/\lambda} e^{-ik_{n_e,2\omega}\zeta} d\zeta \right|^2 \qquad (S22)$$

For perfect Type-I phase matching, $\Delta k = 0$, $n_{o,\omega} = n_{e,2\omega}$ and the exponential term reduces to 1, and the SH intensity is:

$$I_{SH,0}(L,\alpha) \propto L^2 (E_0^{FF})^4 \cos^4\alpha \qquad (S23)$$

In the dynamic case, the principal axes of the BBO crystal are rotated, however, the $\chi^{(2)}$ tensor remains unchanged. This is confirmed by DFPT calculations, which show that even for relatively large phonon amplitudes the variation of the $\chi^{(2)}$ tensor elements remains negligible with respect to the static $d_{22}$ value (see Supplementary Fig. 6). Therefore, the relevant components of the input field remain the x and y-components in the laboratory frame.
Again, the linearly polarized input field can be expressed as:

$$E^{FF} = E_0^{FF} (\sin\alpha, \cos\alpha) \qquad (S24)$$

However, now, upon THz excitation the principal axes are rotated by an amount $\gamma$. To calculate the propagation phase-advance, the components are rotated onto the new principal axes coordinate system by $R_\gamma$, before applying the linear phase retardation and then rotated back to the original lab-frame coordinate system to find the FF components as a function of the propagation distance $\zeta$. In this case the fundamental field components are given by the expression:

$$E^{FF}(\zeta,\gamma) = R_\gamma^T \Gamma(\zeta) R_\gamma E^{FF} \qquad (S25)$$

Here, $R_\gamma$ is the rotation matrix associated with the rotation of the ordinary and extraordinary axes (see Fig. 4a of the main text), and is written as:

$$R_\gamma = \begin{pmatrix} \cos\gamma & -\sin\gamma \\ \sin\gamma & \cos\gamma \end{pmatrix}, \qquad (S26)$$

and $R_\gamma^T$ is its transpose.
If the rotation of the principal axes is considered to be constant ($\gamma_0$) and again in the undepleted pump approximation, the intensity of the SH polarized along lab-frame x-axis after propagating to a depth $z$ inside the crystal is:

$$I_{SH,THz}(z,\alpha,\gamma_0) \propto \left| \int_0^z \left(E_y^{FF}(\zeta,\gamma_0)\right)^2 e^{-ik_{n_e,2\omega}\zeta} d\zeta \right|^2. \qquad (S27)$$

Upon substitution, three terms appear in the integral with distinct arguments in the exponential corresponding to Type-0, Type-I and Type-II wave mixing processes. If the Type-0 and Type-II processes are neglected due to a lack of phase matching, i.e. $n_{e,\omega} \neq n_{e,2\omega}$ and $n_{e,\omega} + n_{o,\omega} \neq 2n_{e,2\omega}$, and only the Type-I process is considered, the integral in the intensity term becomes:

$$E_{SH,THz}(z,\alpha,\gamma_0) \propto E_0^2 \cos^2(\alpha - \gamma_0) \frac{2\sin(\Delta k z/2)}{\Delta k} e^{i\Delta k z} \qquad (S28)$$

Where $\Delta k = (\omega/c)(n_{0,\omega} - n_{e,2\omega}) = (\omega/c)\Delta n$.
For small rotations of the principal axes ($\gamma_0 \ll \pi/2$), the expression for $E_{SH,THz}$ can be expanded.

$$E_{SH,THz}(z,\alpha,\gamma_0) \propto E_0^2 z\, \text{sinc}(\Delta k z/2)\, e^{i\Delta k z} \cos\alpha\, [\cos\alpha + 2\gamma_0 \sin\alpha + O[\gamma_0^2]] \qquad (S29)$$

And, finally, the SH intensity to 1$^{\text{st}}$-order in $\gamma_0$ becomes:

$$\begin{aligned} I_{SH,THz}(z,\alpha,\gamma_0) &= |E_{SH,THz}(z,\alpha,\gamma_0)|^2 \\ &\propto E_0^4 z^2 \, \text{sinc}^2(\Delta k z/2) [\cos^4\alpha + 4\gamma_0 \cos^3\alpha \sin\alpha] \end{aligned} \qquad (S30)$$

Assuming perfect Type-I phase matching, $\Delta k = 0$, the intensity of the SH is given by the expression:
$$I_{SH,THz}(z, \alpha, \gamma_0) \propto E_0^4 z^2 \cos^3 \alpha \, (\cos \alpha + 4 \gamma_0 \sin \alpha) \tag{S31}$$

At equilibrium, $\gamma_0 = 0$, and the conventional expression for SH intensity is recovered, in agreement with the angular dependence observed in the experiment (Fig. 3b of the main manuscript).

Based on equation S31, the SH intensity modulation can be approximated as:
$$\Delta I_{SH.THz} = I_{SH,THz} - I_{SH,0} \approx 4\rho E_0^4 z^2 \gamma_0 \cos^3 \alpha \sin \alpha \tag{S32}$$

where $\rho = \frac{\omega^2 d_{NL}^2}{n_\omega^2 c^2}$ is the proportionality constant and, $d_{22}$ the nonlinear coefficient of Type-I SHG[11].

In the dynamical case, for a finite phonon amplitude $Q_0$ and rotation of the principal axes $\gamma_0$, the analytical expression for the SH intensity modulation in Eq. S32 matches the angular dependence observed in the experiment.

In the presence of optical absorption due to IR-active vibrations, the THz electric field undergoes exponential attenuation within the crystal. As a results, the phonon amplitude decreases as a function of the propagation depth $\zeta$ and can be approximated with an exponential decay, $Q(\zeta) = Q_0 \, e^{-\zeta/\delta_{ph}}$, where $\delta_{ph}$ is the phonon propagation depth. Consequently, along the propagation, the parameter $\gamma$, being proportional to $Q$, also undergoes exponential depletion according to $\gamma(\zeta) = \gamma_0 e^{-\zeta/\delta_{ph}}$. Then, including the propagation dependence of $\gamma(\zeta)$, the SH field $E_{SH,THz}$ can be written as:
$$E_{SH,THz}(z, \alpha, \gamma_0, \delta_{ph}) = \int_0^z \left(E_y^F(\zeta, \gamma_0, \delta_{ph})\right)^2 e^{-\frac{i2n_o\pi\zeta}{\lambda/2}} d\zeta \tag{S33}$$

In this case, the expression for the SH intensity modulation for $z = L$ is found to be:
$$\Delta I_{SH,THz} = 4\rho E_0^4 \, L \, \gamma_0 is \, \cos^3 \alpha \sin \alpha \, \Lambda(\delta_{ph}, L). \tag{S34}$$

After substituting the static SH intensity at $\alpha = 0$, $I_{SH,0}(0) = \rho E_0^4 z L^2$, the modulation becomes:
$$\Delta I_{SH,THz}(\alpha) = 4 \frac{I_{SH,0}(0)}{L} \gamma_0 \, \cos^3 \alpha \sin \alpha \, \Lambda(\delta_{ph}, L), \tag{S35}$$
which corresponds to Eq. 3 of the main manuscript.

Equation S35 is similar to Eq. S32, but it includes the depletion factor:
$$\Lambda(\delta_{ph}, z = L)$$
$$= \left(4e^{-\frac{z}{\delta_{ph}}}\right)$$
$$\cdot \left(\frac{-4\pi^2 \Delta n^2 + 4e^{\frac{z}{\delta_{ph}}}\pi^2 \Delta n^2 - \frac{\lambda^2}{\delta_{ph}^2} + \frac{\lambda^2}{\delta_{ph}^2} \cos\left(\frac{2\pi z \Delta n}{\lambda}\right) - \frac{2\pi\Delta n\lambda}{\delta_{ph}} \sin\left(\frac{2\pi z \Delta n}{\lambda}\right)}{\frac{4\pi^2 \Delta n^2}{\delta_{ph}} + \frac{\lambda^2}{\delta_{ph}^3}}\right) \tag{S36}$$

To evaluate the depletion factor, the wavelength-dependent refractive indices for BBO at room temperature were taken from Ref.[12].

Eq. S35 shows that the SHG modulation intensity $\Delta I_{SH,THz}(\alpha)$ scales with the parameter $\gamma_0$, and therefore with the phonon amplitude $Q_0$, as expected and reproduces the same angular dependence observed experimentally. This equation has been used to fit the polarization-resolved differential SHG modulation (see Fig. 4b in the main text) and to extract the value of $\gamma_0$. A key quantity in this analysis is the phonon penetration depth, $\delta_{\text{ph}}$, which is determined to be 30.2 $\mu m$ by finite-difference time-domain (FDTD) simulations (Fig. 4c in the main text).

**7. Higher-order Nonlinear Processes**

Alongside the phonon-mediated electro-optic modulation of the optical axis - a $\chi^{(2)}$ process that modulates the SHG through phase matching – higher-order THz-mediated $\chi^{(3)}$ effects. These $\chi^{(3)}$ processes have been described within the framework of Terahertz-Field–Induced Second Harmonic (TFISH). It should be noted, however, that in previous TFISH works[13–16], either the phonons were not resonantly excited, or critically, the SHG process was phase-mismatched. In this case, the frequency conversion is confined to a coherence length that is typically on the order of microns – commensurate with a few wavelengths of the fundamental beam. Within this short-interaction regime the THz-driven phonons directly modulate the second-order nonlinearity resulting in a $\chi^{(3)}$ effect, while second-order electro-optic variation of the refractive index ellipsoid and its impact on non-existent phase matching conditions are negligible.

In the current experiment, $\chi^{(3)}$ effects are found to have negligible impact on SHG modulation under optimal phase matching conditions, as they cannot reproduce the angular dependence observed in the experiment, the $\chi^{(3)}$ SHG modulation increases and decreases uniformly as a function of time delay for all input polarisations (rather than exhibiting a vice-versa behaviour between neighbouring lobes as observed in the experiment), and finally, the magnitude of the $\chi^{(3)}$-effects on the SHG intensity modulation are much smaller to those observed in the experiment.

To isolate and study the $\chi^{(3)}$-effects, additional experiments were performed in a c-cut ($\theta = 0°$) BBO crystal (MolTech GmbH, 500 µm thick), which is far from Type-I phase-matching conditions (Supplementary Fig. 5a-b). The crystal quality and orientation were verified by Laue x-ray diffraction. In this configuration, the THz and the NIR pulses propagate along the optic axis (c-axis) and the refractive index is $n_o$ along $x$- and $y$-axes in the laboratory frame. Therefore, the crystal is isotropic and the coherence length is identical for all input polarization states.

In the phase-mismatched case, the equilibrium SHG conversion is two orders of magnitude lower than in the phase-matched case and the nonlinear coefficient $d_{22}$ dominates the SHG conversion. Analytically, the SH intensity along the lab-frame $x$-axis is, $I_{SHG,x} \propto |d_{22}E_x^2 - d_{22}E_y^2|^2$, where $E_x = E_0^{FF} \sin\alpha$ and $E_y = E_0^{FF} \cos\alpha$ are the projections of fundamental field $E^{FF}$ along the lab-frame $x$- and $y$-axes. From the expression, a 4-fold symmetry in input polarisation angle $\alpha$ is expected, with local maxima on the x- and y-axes. Indeed, at equilibrium in the phase-mismatched case, the observed SHG has a polarization dependence in agreement with the analytical expression (Fig S5c).

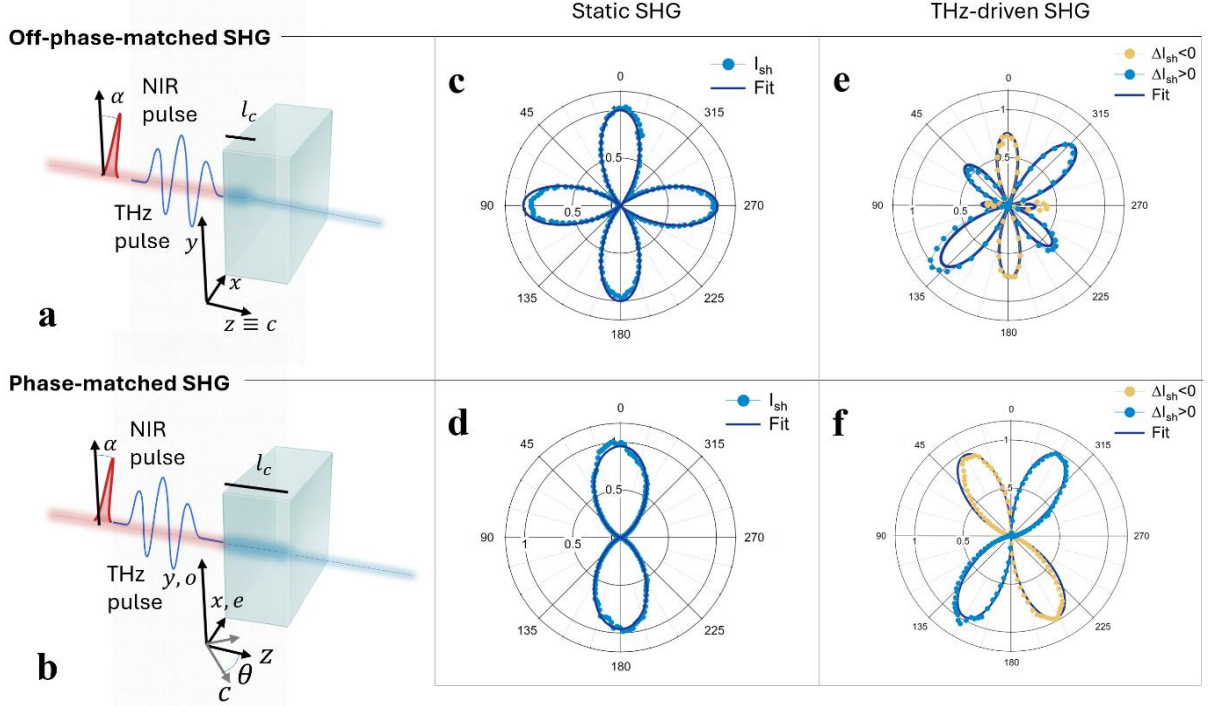

***Supplementary Fig. 5. Isolating the TFISH contribution to THz-driven SHG modulation. a**, Off–phase-matched geometry: the NIR and THz pulses propagate along the crystal c axis; SHG is collected along y. **b**, Phase-matched SHG geometry. **c**, Static "off-phase-matched" SHG: experimental data and fit curve with the function: $I_{SHG} \propto |d_{22}E_x^2 - d_{22}E_y^2|^2$. **d**, Static phase-matched SHG: experimental data with fit: $I_{SHG} \propto |d_{22}E_y^2|^2$, where the $E_x^2$ is not present because not phase-matched. **e**, THz-induced SHG modulation in the off–phase-matched case: data and fit: $\Delta I_{SHG} \propto a_1 E_x^4 + a_2 E_y^4 + a_3 E_x^2 E_y^2 + a_4 E_x^3 E_y + a_5 E_x E_y^3$, modelling TFISH-driven modulation of the $\chi^{(2)}$ tensor components, see Ref.[14]. The $a_3$ is dominant and produces lobes oriented at 45°; $a_4$ and $a_5$ are over one order of magnitude smaller. The maximum SHG intensity modulation is below 1% at the maximum THz field. **f**, THz-induced SHG modulation under phase-matched conditions: the angular symmetry differs qualitatively from panel **e** and cannot be described by TFISH alone.*

In contrast, when equilibrium SHG measurements are performed under optimal Type-I phase matching conditions, only the SH due to the Type-I process (SH along the extraordinary in the lab frame, x-axis) is generated by the component of the FF field projected along the ordinary, y-axis ($o + o \rightarrow e$). Contributions from the other input fundamental field components are negligible due to a lack of phase matching. Analytically, this results in a static SHG signal $I_{SHG} \propto |d_{22}E_x^2|^2 \propto \cos^4 \alpha$ with 2-fold angular symmetry (see Supplementary Fig. 5d).

In the dynamical case, the THz excitation modulates the intensity of the SH. The effect can be considered from two perspectives, either the THz field can be viewed to modulate the 2$^{nd}$-order nonlinear susceptibility such that $\chi^{(2)} \rightarrow \chi^{(2)}(t)$, or the effect can be seen to occur through a 4-wave mixing process involving the 3$^{rd}$-order nonlinear susceptibility $\chi_0^{(3)}$ and the THz and fundamental fields. The second scenario is commonly used in the description of TFISH experiments and is why THz-Field Induced Second Harmonic is often known as a 3$^{rd}$-order $\chi^3$-effect. Within the TFISH framework, the intensity of the SHG signal has three terms, a time-dependent heterodyne term which depends linearly on the THz electric field, a homodyne term and a DC offset term[17]:

$$I_{SHG}(t) \propto \left|\left(\chi^{(2)} + \chi^{(3)} E_{THz}(t)\right) E_{FF}^2\right|^2$$
$$\propto \underbrace{\left|\chi^{(2)} E_{FF}^2\right|^2}_{DC} + \underbrace{2\text{Re}\left(\chi^{(2)}\chi^{(3)} E_{THz}(t) E_{FF}^4\right)}_{Heterodyne \propto E_{THz}(t)} \quad \text{(S37)}$$
$$+ \underbrace{\left|\chi^{(3)} E_{THz}(t) E_{FF}^2\right|^2}_{Homodyne \propto |E_{THz}(t)|^2}.$$

In the alternative perspective, where the THz electric field is assumed to modulate the 2$^{nd}$-order susceptibility tensor, $\chi^{(2)}(t)$ can be written as:

$$\chi^{(2)}(t) = \chi^{(2)} + \Delta\chi^{(2)}(t) \text{ and } I_{SHG}(t) \propto \left|\chi^{(2)}(t)E_{FF}^2\right|^2, \tag{S38}$$

where $\chi^{(2)}$ is the static nonlinear susceptibility tensor and $\Delta\chi$ is the THz induced, time-dependent modulation.

The pictures are equivalent, as the time-dependent change in the nonlinear susceptibility is assumed proportional to the THz field, $\Delta\chi^{(2)}(t) \sim \chi^{(3)}E_{THz}(t) \sim \chi^{(3)}Q(t)$.

If only a single input FF polarization is used, the heterodyne term in the time-dependent intensity might erroneously be considered as the source of the SH modulation in the phase-matched experiment. The effect is linearly dependent in strength on the THz electric field, it is phase-locked to the THz electric field, and it has a frequency dependence that matches the THz-driven phonon. However, the strength of the 3$^{rd}$-order TFISH effect is expected to be much weaker than the 2$^{nd}$-order effect, and perhaps even more conclusively, when the angular symmetry of the intensity modulation is included in the analysis, it can be shown that 3$^{rd}$-order processes cannot be the underlying source of the observed dynamics in the phase-matched case.

To investigate the polarization dependence, the perspective employing a time-dependent $\chi^2(t)$ will be used for analysis and interpretation of the dynamical SH signal in both the phase-matched and phase-mismatched geometries. By convention, the nonlinear susceptibility is expressed in terms of the nonlinear coefficient $d = \frac{1}{2}\chi^{(2)}$, so that the dynamical SH field amplitude is:

$$E_{SH,THz}^{xtal} \propto (d_0 + \Delta d)E_{xtal}^{FF}E_{xtal}^{FF}, \tag{S39}$$

where $E_{xtal}^{FF}$ is the fundamental field in the crystal-frame, $d_0$ is the nonlinear coefficient at the equilibrium and $\Delta d$ is the modulation induced by the THz field.

By applying a coordinate transformation from the lab-frame to crystal-frame, evaluating the SH fields that are generated in the BBO and then transforming back to the lab-frame, the resultant SH field components can be related to the input polarisation of the fundamental field in the lab frame, which is defined by $\alpha$,

$$E_{SH,THz}^{lab}(\alpha) = R^T(d_0 + \Delta d)f_{3\to 6}\left(R\left(E_{FF}^{lab}(\alpha)\right)\right). \tag{S40}$$

Here, $f_{3\to 6}$ is the operator that maps vector components $(x', y', z')$ of the input field onto the $(x'^2, y'^2, z'^2, 2y'z', 2x'z', 2x'y')$ terms corresponding to the square of the input field. The components of the fundamental field in the lab frame are, $E_0^{FF}(\sin\alpha, \cos\alpha, 0)$, with $\alpha$ the polarization angle with respect to the $y$-axis. $R$ is the rotation matrix that maps the field components from the lab frame to the crystal frame and is a function of the crystal geometry ($\theta = 29.3, \phi = 90$).

Finally, the SH intensity as a function of input polarisation can be written as:

$$I_{SH,THz}(\alpha) \propto \left|E_{SH,THz}^{lab}(\alpha)\right|^2 \propto \left|R^T(d_0 + \Delta d)f_{3\to 6}\left(R\left(E_{FF}^{lab}(\alpha)\right)\right)\right|^2 \tag{S41}$$

The output intensity in the dynamical case contains interference terms involving the static 2$^{nd}$-order nonlinearity and time-dependent 2$^{nd}$-order nonlinearity – or equivalently the $\chi^{(3)}$ nonlinear susceptibility.

When the $x$-component in the lab frame is isolated with a polariser (corresponding to the extraordinary axis in the phase matched geometry), and for generality if all elements in the $\Delta d$ tensor are allowed to be non-zero, the polarisation dependent intensity difference includes the five following terms ($E_z = 0$):

$$\Delta I_{SH,THz}^x(\alpha) = I_{SH,THz}^x - I_{SH,0}^x \tag{S42}$$
$$\propto (a_1)E_x^4 + (a_2)E_y^4 + (a_3)E_x^2E_y^2 + (a_4)E_x^3E_y + (a_5)E_xE_y^3,$$

where:
$$E_x = E_0^{FF} \sin \alpha \quad ; \quad E_y = E_0^{FF} \cos \alpha. \tag{S43}$$

In Eq. S42, the $a_1, \ldots, a_5$ coefficient depends on the static and modulated nonlinear tensor elements. Given the initial symmetry of the BBO crystal ($d_{15}, d_{22}, d_{33} \neq 0$), the THz induced variation in the intensity of the second harmonic can have an angular dependence with terms proportional to, $\cos^4 \alpha, \sin^4 \alpha, \cos^2 \alpha \sin^2 \alpha, \cos \alpha \sin^3 \alpha$ and $\cos^3 \alpha \sin \alpha$, and the proportionality will depend to 1st or 2nd order on the time-dependent second order nonlinearity, or the 3rd- order susceptibility $\chi^{(3)}$.

The angular dependence of the SH intensity modulation in the phase-mismatched case is shown in Supplementary Fig. 5e. Here, as previously stated, the TFISH, $\chi^{(3)}$-effects are isolated and the angular dependence can be fit with an equation of the form in S42. The dominant fit parameters are found to be associated with the $\cos^4 \alpha, \sin^4 \alpha$ terms, which scale the lobes on the $x$- and $y$-axes, and the $\cos^2 \alpha \sin^2 \alpha$ term, which scales the lobes along the diagonals. Small adjustments to the rotation of the polar fit can be made using the remaining terms. In the time-domain, the onset of THz excitation results in a uniform decrease of the SHG signal on the axes, $\Delta I_{SHG} < 0$, which is consistent with a THz-induced decrease of the $d_{22}$ coefficient. At the same time, the onset of THz excitation results in a uniform increase of the SHG signal on the diagonals, $\Delta I_{SHG} > 0$.

The angular dependence in the Type-I phase-matched case is shown in Supplementary Fig. 5f. Here, in contrast to the phase-mismatched case, the strongest variation in the SH intensity is found at $\pm 30°$ and the signal on the axes is minimised. Moreover, the sign of the effect alternates between lobes such that if one lobe experiences an increase in SH intensity, the neighbouring lobe exhibits a decrease. If the dynamics were explained by TFISH or, equivalently, $\chi^{(3)}$ effects, one would expect that the $\cos^4 \alpha, \sin^4 \alpha$ and $\cos^2 \alpha \sin^2 \alpha$ terms would all play a role in the SH modulation, yet there is no clear signature of any of these terms in the angular dependence of the phase-matched measurements. This is especially notable for the $\sin^4 \alpha$ term, which involves a heterodyne term that mixes two Type-I phase matched processes, and would be expected to have the strongest response.

Instead, only the $\cos^3 \alpha \sin \alpha$ term from the TFISH scenario is present in the phase-matched angular dependence. A closer look at the elements in the proportionality coefficient of the $\cos^3 \alpha \sin \alpha$ term, reveals component that is linear in THz field strength and scales with $O(\Delta d)$ or $O\left(\chi_0^{(3)}\right)$, and a component that is quadratic in THz field strength and scales with $O^2(\Delta d)$ or $O^2\left(\chi_0^{(3)}\right)$.

The homodyne contribution can be excluded because the SH intensity modulation varies linearly with the THz field strength (Fig. 2c of the main manuscript). Furthermore, as the homodyne contribution is proportional to $\sim Q^2$, it would need to be driven by a phonon at half the observed modulation frequency. However, in the experiment, when the driving THz was filtered to excite only a phonon at 2.15THz, the intensity modulation is nearly extinguished (Fig. 2, main manuscript).

There are also reasons to suggest that the heterodyne term does not contribute significantly to the phase-matched intensity modulation either. Importantly, the heterodyne interference term involves an SH component proportional to $d$ that results from the Type-I phase-matched process, but the other component, proportional to the $\Delta d$ time-dependent nonlinearity, results from a Type-II process, which is NOT phase matched.

In summary, the angular dependence in the phase-matched case does not show the features one would expect from the phase-mismatched reference measurements, and the features that are expressed do not appear to have originated from a $\chi^{(3)}$ process.

The features in the phase-matched case are however fit perfectly by the cascaded 2nd-order model involving electro-optic modification of the index of refraction ellipsoid and its effect on phase-matching in the usual SHG process with static 2nd-order nonlinearity.

## 8. Numerical estimate of the $\chi^{(3)}$ contribution in the phase-matched case

Qualitative analysis of the angular symmetry in the phase-matched experiment and comparison with the symmetry observed in the phase-mismatched experiments fails to provide evidence of $\chi^{(3)}$ or TFISH effects as the definitive source of the THz-driven SH modulation in BBO. However, higher order effects are always present, and additional insight can be gained by a more quantitative evaluation of the strength of these effects.

In the TFISH framework, the intensity of the SHG signal is proportional to three terms, a time-dependent heterodyne term, a time dependent homodyne term and a DC offset term:

$$I_{SHG}(t) \propto \left|\left(\chi^{(2)} + \chi^{(3)} E_{THz}(t)\right) E_{FF}^2\right|^2$$
$$\propto \underbrace{\left|\chi^{(2)} E_{FF}^2\right|^2}_{DC} + \underbrace{2Re(\chi^{(2)}\chi^{(3)} E_{THz}(t) E_{FF}^4)}_{Heterodyne \propto E_{THz}(t)} \quad (S44)$$
$$+ \underbrace{\left|\chi^{(3)} E_{THz}(t) E_{FF}^2\right|^2}_{Homodyne \propto |E_{THz}(t)|^2}$$

To link the intensity modulation to the optically driven phonons in the BBO, according to Ref.[18], $\chi^{(3)} E_{THz}(t)$ is identical to a time-dependent perturbation of the second-order non-linearity and can be rewritten in terms of the effective nonlinear coefficient $d_{eff}$ and the phonon displacement $Q$:

$$\chi^{(3)} E_{THz}(t) = 2\frac{\partial d_{eff}}{\partial Q} Q(t) \rightarrow \Delta d = \frac{\partial d_{eff}}{\partial Q} Q \quad (S45)$$

Therefore, the heterodyne contribution is proportional to the amplitude of the phonon, which is assumed proportional to the THz field.

DFT calculations of the phonon displacements can then be used to estimate the time-dependent 2nd-order nonlinear coefficients, $\Delta d$. Calculations of the phonon-amplitude dependent nonlinear coefficients are shown in Fig S6. By combining these coefficients with the phonon penetration depth $\delta_{ph}$ (calculated independently by DFPT), a simple two-layer model of the BBO crystal can be used to estimate the strength of the 3rd-order effects. This model consists of a uniformly pumped layer with a constant phonon amplitude $Q_0$ that is $\delta_{ph}$ thick with the additional time-dependent nonlinearities, while the remainder of the crystal is considered to be at equilibrium.

The Type-I phase-matched component, which is proportional to $d_{22}^2 |E_y^{FF}|^4$ will be affected by the time-dependent term $\Delta d_{22}$, which is expected to have a maximum value of $\Delta d_{22} = 0.1\ pm/V$ with $d_{22} = 2.9\ pm/V$. In the 2-layer model which is expected to overestimate the strength of the effect, the relative 3rd-order intensity modulation of this component is calculated to be 0.68%, far below the modulation observed in the experiment. Furthermore, in terms of the angular dependence, this modulation will result in a uniform scaling of the original equilibrium 2-fold symmetric pattern, which is not observed in the experiment.

The strength of the Type-I/Type-II heterodyne component can also be calculated using the other time-dependent nonlinear components, $\Delta d_{25}$, $\Delta d_{26}$, and $\Delta d_{35}$. With the same 2-layer model, the relative intensity modulation that can be expected from this term is approximately 0.3%. While this value is again expected to be an overestimate, it is still far below the modulation observed in the experiment.

Again, the estimates of the strength of the $\chi^{(3)}$-effect fail to support a TFISH explanation of the intensity modulation in the phase-matched experiment.

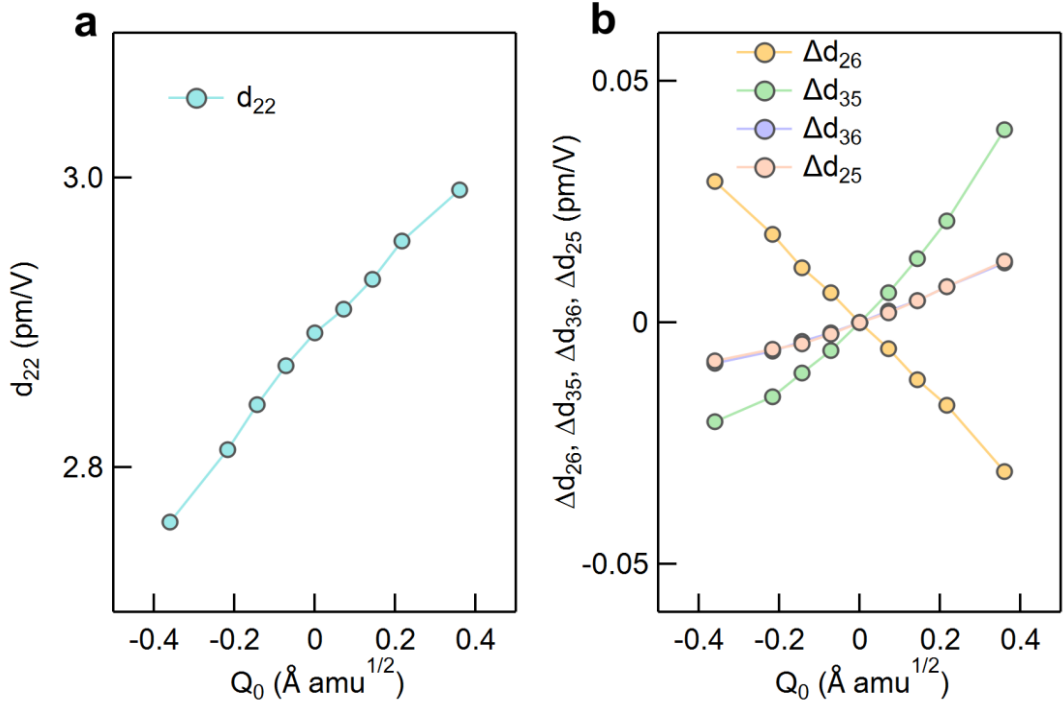

**Supplementary Fig. 6. a**, *DFT calculated modulation of the nonlinear coefficients $d_{22}$ as a function of lattice displacement $Q_0$. **b**, Modulation of the nonlinear coefficients $\Delta d_{26}$, $\Delta d_{35}$, $\Delta d_{25}$, and $\Delta d_{36}$ as a function of $Q_0$. At the equilibrium structure, these coefficients are approximately zero.*

## 9. Estimation of the vibrational electro-optic coefficient and its frequency dependence

This section describes in detail the calculation of the vibrational electro-optic coefficient magnitude near the phonon resonance, as obtained from DFPT and from experimentally estimated values.

**Calculation of the EO coefficient from DFPT:** The modulation of the refractive index $\Delta n$ is related to the phonon coordinate $Q$ by:

$$\Delta n = \frac{n_x^3 \prime r_{22} Q}{2}, \qquad \text{Eq. S46}$$

see Eq.2 in the main manuscript. Here, $r_{22}$ denotes the phonon–optic coupling coefficient that quantifies the coupling between the phonon amplitude and the resulting refractive-index modulation.

From the DFTP results, $\Delta n$ varies linearly with $Q$ by which a value of $r_{22} = 0.0049$ Å$^{-1}$ $(amu)^{1/2}$ can be extrapolated from a linear fit, as discussed in the main text.

The standard vibrational EO coefficient, which is denoted here as $r_{22}^E$ describes the variation of the index of refraction with respect to the applied electric field strength. This coefficient can be given in SI units $(pm/V)$.

The conventional linear relationship between the phonon coordinate $Q$ and the electric field $E_{THz}$: $Q(\omega) = \frac{\sqrt{\varepsilon_0 V}\omega_p}{(\omega_Q^2 - \omega^2 - i\Gamma\omega)} E_{THz}(\omega)$ is required to convert between the two coefficients. Substituting this relationship into Eq. S46 yields

$$\Delta n = \frac{n_x^3}{2} r_{22} \frac{\sqrt{\epsilon_0 V}\omega_p}{(\omega_Q^2 - \omega^2 - i\Gamma\omega)} E_{THz} = \frac{n_x^3}{2} \tilde{r}_{22}^E E_{THz} \qquad \text{Eq. S47}$$

Here, $V$ is the volume of the unit cell, $\epsilon_0$ is the vacuum permittivity, $\omega_Q$, $\omega_p$ and $\Gamma$ are the phonon frequency, phonon plasma frequency and the linewidth, respectively. Here, $\tilde{r}_{22}^E$ is the complex EO coefficient in SI units, and taking its real part:

$$r_{22}^E = Re\{\tilde{r}_{22}^{EO}\} = r_{22}\frac{\sqrt{\epsilon_0 V}\omega_p(\omega_Q^2-\omega^2)}{\left(\omega_Q^2-\omega^2\right)^2-(\Gamma\omega)^2} = r_{22}^E(0)\frac{(\omega_Q^2-\omega^2)}{\left(\omega_Q^2-\omega^2\right)^2-(\Gamma\omega)^2}. \qquad \text{Eq. S48}$$

Note that Eq. S48 corresponds to the usual expression of the vibrational EO coefficient (see e.g. Ref. [19]), also shown in in Eq. S47.

Using $r_{22} = 0.0049$ Å$^{-1}$ $(amu)^{\frac{1}{2}}$ from the DFPT-derived refractive-index modulation, and $\omega_P = 6.57$ THz, $\omega_Q = 4.32$ THz and $\Gamma = 0.19$ THz determined by infrared spectroscopy, a maximum positive value of $r_{22-max}^E = 55\frac{pm}{V}$ at $\omega_* = \sqrt{\omega_Q^2 - \Gamma\omega_Q} = 4.25$ THz is obtained. As discussed in the main text, the EO coefficient does not peak exactly at the phonon frequency $\omega_Q$, the real part attains its maximum just below resonance at $\omega_*$ which is also consistent with the frequency peak of the SHG modulation response (see Fig. 5b of the main text).

A comparison of the calculated EO coefficient with experimental values can be performed in the DC limit. Extrapolating Eq. S48 to $\omega \to 0$ yields $r_{22}^E(0) = 6.6$ pm/V, which is larger than the commonly reported experimental DC value of approximately 2.5 pm/V for BBO (Ref. [20]). This discrepancy likely reflects uncertainties in the experimental determination of the phonon plasma frequency, as well as the limitations of the single-phonon DFPT approach in accurately describing the low-frequency electro-optic response, which includes both electronic contributions and the cumulative effect of multiple phonon modes.

**Experimental estimation of the EO coefficient from broadband THz pumping:** Experimentally, the value of $r_{22}^E$ can be estimated from the measured SHG modulation dynamics. As discussed in the main text, the relative SHG modulation, at fixed polarisation angle $\alpha$ is: $\Delta I_{rel}(t,\alpha) = \Delta I_{SH}(t,\alpha)/I_{SH,0}(\alpha)$, where $I_{SH,0}(\alpha) = I_{SH,0}(0)\cos^4\alpha$. Measurements made at $\alpha = 30°$ were used to evaluate the EO coefficient.

Analytically, the modulation of the SH dynamics $\Delta I_{SH}(\alpha,t)$ in the time domain as a function of $\alpha$ (Eq. 2 in the main text) is:

$$\Delta I_{SH}(\alpha,t) = 4\,I_{SH,0}(0)\gamma(t)\frac{\Lambda(\delta_{ph},L)}{L}\cos^3\alpha\sin\alpha, \qquad \text{Eq. S49}$$

therefore,

$$\Delta I_{rel}(\alpha,t) = 4\,\gamma(t)\frac{\Lambda(\delta_{ph},L)}{L}\frac{\sin\alpha}{\cos\alpha} = 4\kappa(\theta)r_{22}Q(t)\frac{\Lambda(\delta_{ph},L)}{L}\tan\alpha. \qquad \text{Eq. S50}$$

Here, $\kappa(\theta = 29.3°) = 65$ is the crystal geometrical factor, $\Lambda$ is the parameter that accounts for the finite phonon penetration depth $\delta_{ph} = 30.2$ μm, within a crystal of thickness $L = 300$ μm and $Q(t)$ is the time dependent phonon coordinate. $Q(t)$ can be calculated using the phonon equation of motion leaving $r_{22}$ as the only free parameter to find agreement with the experimentally measured relative modulation.

The phonon dynamics are described by:

$$\ddot{Q} + \Gamma\dot{Q} + \omega_Q^2 Q = \sqrt{\epsilon_0 V}\omega_p\tilde{E}_{THz}(t), \qquad \text{Eq. S51}$$

where $\epsilon_0$ is the vacuum permittivity, $V = 596$ Å$^3$ is the unit cell volume, and $\omega_Q = 4.32$ THz, $\omega_p = 6.57$ THz and $\Gamma = 0.19$ THz are the phonon frequency, phonon plasma frequency and the linewidth measured by THz-FTIR spectroscopy. $\tilde{E}_{THz}(t)$ is the THz field transmitted through the air-sample interface: $\tilde{E}_{THz}(t) = \frac{1}{2\pi \int \tilde{t}(\omega) E_{THz}(\omega) e^{-i\omega t} d\omega}$, where $\tilde{t}(\omega)$ is the complex Fresnel transmission calculated from the BBO refractive index (determined by THz FTIR), $E_{THz}(\omega)$ is the spectrum of the THz pump field measured by electro-optic sampling (EOS).

Solving Eq. S51 for $Q(t)$ and substituting into Eq. S50, agreement between the calculated relative modulation and measurement is obtained for a phonon-optic coupling coefficient, $r_{22} = 0.0130$ Å$^{-1}$ (amu)$^{1/2}$ (Fig S7) with corresponding EO coefficient, $r^E_{22-max} = 145 \frac{pm}{V}$.

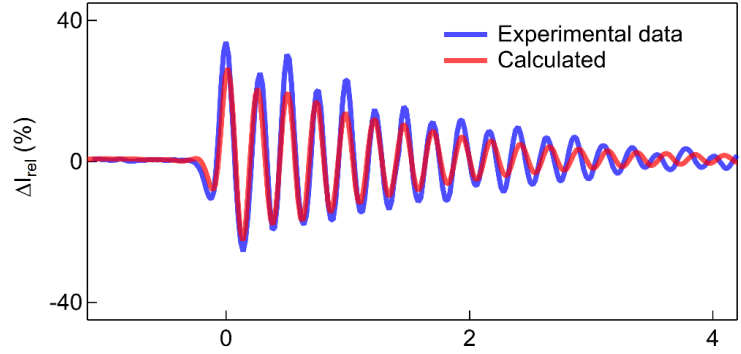

***Supplementary Fig. 7.*** *Experimentally measured $\Delta I_{rel}$ and calculated one using Eq. S50.*

**Estimation of EO coefficient from narrowband THz pumping:** Broadband THz pumping contains spectral components far from the phonon resonance which may interfere with the determination of $r^E_{22}$. Therefore, an important cross-check is to evaluate $r^E_{22}$ with a narrowband THz excitation using band-pass THz filters to isolate spectral components centered around the phonon frequency.

The optical setup was based on a 1-kHz repetition-rate Ti:sapphire laser. THz radiation was generated in a DSTM crystal and driven by OPA signal pulses with an energy of 0.44 mJ per pulse. The resulting THz waveform and the corresponding field-strength estimation were obtained via EOS using a 200-µm-thick GaP crystal. The THz electric field was determined using the conventional expression (see, e.g., Ref.[21]), including the Fresnel transmission coefficient. The measured temporal waveform and its Fourier amplitude spectrum are shown in Supplementary Fig. 8a and Supplementary Fig. 8b, respectively.

The relative SH modulation dynamics $\Delta I_{\text{rel}}(\alpha, t)$, were recorded at the polarization angle ($\alpha = 30°$), and the corresponding Fourier amplitude spectrum is shown in Supplementary Fig. 8c and Supplementary Fig. 8d, respectively.

Following the same procedure as in the broadband THz case, the phonon optic coupling coefficient was found to be $r_{22} = 0.0160$ Å$^{-1}$ $(amu)^{1/2}$, which corresponds to $r^E_{22-max} = 178 \frac{pm}{V}$, comparable to the value obtained using the broadband THz excitation.

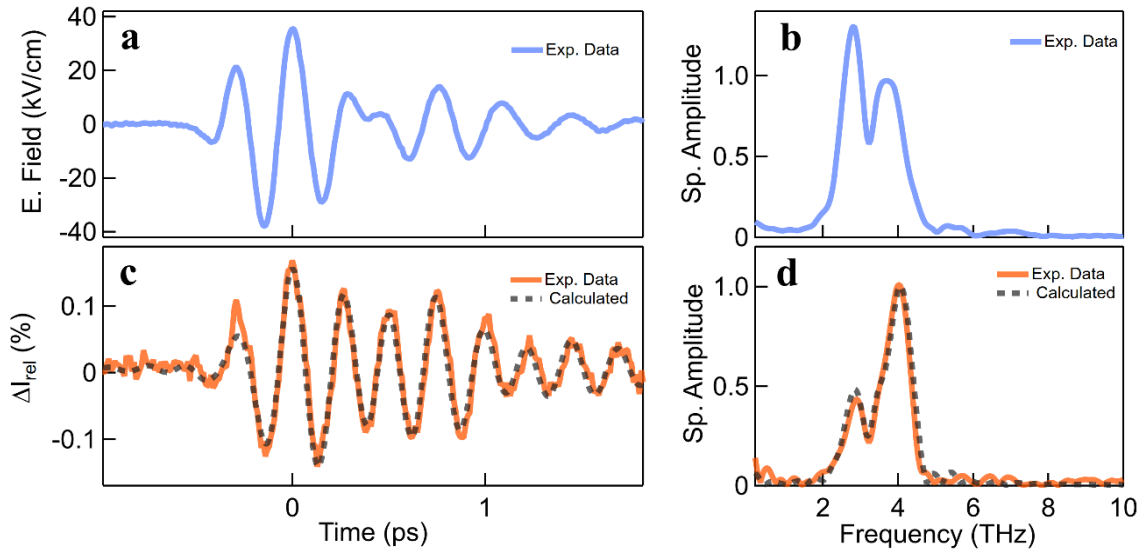

*Supplementary Fig. 8. SHG modulation with a narrowband THz pulse. **a**, Time-domain THz pump waveform measured by EOS. **b**, Amplitude spectra corresponding to **a**. **c**, THz-driven $\Delta I_{rel}$ and calculated one using Eq. S50. **d**, Amplitude spectra corresponding to curves in **c**.*